\documentclass[11pt,a4paper]{article}
\pdfoutput=1
\usepackage{jcappub}
\usepackage{bm}
\usepackage{indentfirst}
\usepackage{amsmath}
\usepackage{graphicx}
\usepackage{float}
\usepackage{amssymb}
\usepackage{subfigure}
\usepackage{amssymb}
\usepackage{psfrag}
\usepackage{hyperref}
\usepackage{tensor}

\usepackage{color}
\usepackage{comment}
\usepackage{CJK}
\usepackage{array}
\usepackage{indentfirst}
\newcommand{\Rin}{R^{\textup{in}}_{l m \omega}} % in mode Teukolsky equation in normalized units
\newcommand{\Rup}{R^{\textup{up}}_{l m\omega}} %up mode Teukolsky equation in normalized units
\newcommand{\Xin}{X^{\textup{in}}_{l m\omega}} % in mode Sasaki-Nakamura equation in normalized units
\newcommand{\Xup}{X^{\textup{up}}_{l m \omega}} %up mode Sasaki-Nakamura equation in normalized units

\begin{document}
\title{Detecting vector charge with extreme mass ratio inspirals onto Kerr black holes}
\author[a]{Chao Zhang,}
\author[b]{Hong Guo,}
\author[c,1]{Yungui Gong\note{Corresponding author.},}
\author[d,e,1]{and Bin Wang}

\affiliation[a]{School of Physics and Astronomy, Shanghai Jiao Tong University, 800 Dongchuan Rd, Shanghai 200240, China}
\affiliation[b]{Shanghai Frontier Research Center for Gravitational Wave Detection, Shanghai Jiao Tong
University, 800 Dongchuan Rd, Shanghai 200240, China}
\affiliation[c]{School of Physics, Huazhong University of Science and Technology, 1037 LuoYu Rd, Wuhan, Hubei 430074, China}
\affiliation[d]{Center for Gravitation and Cosmology, Yangzhou University, 88 South Daxue Rd, Yangzhou, Jiangsu 225009, China.}
\affiliation[e]{School of Aeronautics and Astronautics, 800 Dongchuan Rd, Shanghai 200240, China}

\emailAdd{zhangchao666@sjtu.edu.cn}
\emailAdd{gh710105@gmail.com}
\emailAdd{yggong@hust.edu.cn}
\emailAdd{wang\_b@sjtu.edu.cn}

\keywords{vector charge, gravitational waves, extreme mass ratio inspirals, Kerr black holes}
\abstract{
Extreme mass ratio inspirals (EMRIs) are excellent sources for space-based observatories to explore the properties of black holes and test no-hair theorems.
We consider EMRIs with a charged compact object inspiralling onto a Kerr black hole in quasi-circular orbits.
Using the Teukolsky and generalized Sasaki-Nakamura formalisms for the gravitational and vector perturbations about a Kerr black hole,
we numerically calculate the energy fluxes for both gravitational and vector perturbations induced by a charged particle moving in equatorial circular orbits.
With one-year observations of EMRIs,
we apply the Fisher information matrix method
to estimate the charge
uncertainty detected by space-based gravitational wave detectors such as the Laser Interferometer Space Antenna, TianQin, and Taiji,
and we find that it is possible to detect vector charges as small as $q\sim 0.0049$.
The results show that EMRIs composed of a Kerr black hole with a higher spin $a$ and lighter mass $M$, and a secondary charged object with more vector charge give smaller relative error on the charge, thus constrain the charge better.
The positive spin of the Kerr black hole can decrease the charge uncertainty by about one or two orders of magnitude.}

\arxivnumber{2301.05915}

\maketitle

\section{Introduction}
In 2015 the Laser Interferometer Gravitational-Wave Observatory (LIGO) Scientific Collaboration and the Virgo Collaboration \cite{Abbott:2016blz,TheLIGOScientific:2016agk} directly observed the first gravitational wave (GW) event GW150914 coming from the coalescence of binary black holes (BHs),
and the discovery opened us a new window to understand the property of BHs and gravity in the nonlinear and strong field regimes.
Until now tens of GW events in the frequency range of tens to hundreds Hertz have been confirmed \cite{LIGOScientific:2018mvr,LIGOScientific:2020ibl,LIGOScientific:2021usb,LIGOScientific:2021djp,LIGOScientific:2017vwq,LIGOScientific:2020aai,LIGOScientific:2021qlt}.
However, due to the seismic noise and gravity gradient noise, the ground-based GW observatories can only measure transient GWs in the frequency range $10-10^3$ Hz, radiated by the coalescences of stellar-mass compact binaries.
Apart from transient GW sources, the future space-based GW detectors like the Laser Interferometer Space Antenna (LISA) \cite{Danzmann:1997hm,LISA:2017pwj}, TianQin \cite{Luo:2015ght} and Taiji \cite{Hu:2017mde} will help us uncover unprecedented information about GW sources and fundamental physics \cite{Gong:2021gvw,TianQin:2020hid,Ruan:2018tsw,LISA:2022yao,LISA:2022kgy}.
One of the most conspicuous sources for the future space-based GW detectors is the extreme mass-ratio inspirals (EMRIs) \cite{Amaro-Seoane:2007osp,Babak:2017tow}.
EMRIs, which consist of a stellar-mass compact object (secondary object) with mass $m_p\sim1-100~M_{\odot}$ such as BHs, neutron stars, white dwarfs, etc. orbiting around a supermassive black hole (SMBH) (primary object) with mass $M\sim10^5-10^7~M_{\odot}$,
with the mass ratio $m_p/M$ in the range of $10^{-7}-10^{-4}$,
radiate millihertz GWs expected to be observed by the future space-based GW detectors.

% EMRIs can provide a unique wealth of information about the masses, spins, electric charges, the strong-field physics in the vicinity of BHs, and the astrophysics of their stellar environments, etc \cite{Amaro-Seoane:2007osp,Babak:2017tow,Berry:2019wgg,Fan:2020zhy,Zi:2021pdp,Destounis:2020kss,Destounis:2021mqv,Destounis:2021rko,Cardoso:2021wlq}.
Future detections of EMRIs with space-based detectors can provide highly precise measurements on source parameters such as the BH masses, spins, etc.
In \cite{Barack:2003fp}, the authors introduced a family of approximate waveforms for EMRIs to make parameter estimation with LISA.
For a typical source of $m_p=10~M_{\odot}$ and $M=10^6~M_{\odot}$ with a signal-to-noise ratio (SNR) of $30$,
LISA can determine the masses of both primary and secondary objects to within a fractional error of $\sim 10^{-4}$, measure the spin of the primary object to within $\sim 10^{-4}$, and localize the source on the sky within $\sim 10^{-3}$ steradians.
The improved augmented analytic kludge model \cite{Chua:2017ujo} provides more accurate and efficient GW waveforms to improve the errors of parameters by one order of magnitude.
Thus, EMRIs can be used to precisely measure the slope of the black-hole mass function  \cite{Gair:2010yu} or as standard sirens \cite{Holz:2005df} to  constrain cosmological parameters and investigate the expansion history of the Universe \cite{MacLeod:2007jd,Laghi:2021pqk,LISACosmologyWorkingGroup:2022jok}.
The observations of EMRIs can also help us explore gravitational physics.
For example, they can be used to figure out the spacetime structure around the central SMBH to high precision,
allowing us to test if the spacetime geometry is described by general relativity or an alternative theory
and analyze the environments such as dark matter surrounding the central SMBH \cite{Amaro-Seoane:2007osp,eLISA:2013xep,Eda:2013gg,Eda:2014kra,Barausse:2014tra,Yue:2017iwc,Yue:2018vtk,Babak:2017tow,Berry:2019wgg,Hannuksela:2019vip,Destounis:2020kss,Burton:2020wnj,Torres:2020fye,Barausse:2020rsu,Cardoso:2019rou,Maselli:2020zgv,Maselli:2021men,Guo:2022euk,Zhang:2022rfr,Barsanti:2022ana,Barsanti:2022vvl,Cardoso:2021wlq,Dai:2021olt,Jiang:2021htl,Zhang:2022hbt,Gao:2022hho,Gao:2022hsn,Destounis:2022obl,Liang:2022gdk}.
In \cite{Cardoso:2020iji,Liu:2020vsy,Liu:2020bag},
the authors investigated the eccentricity and orbital evolution
 of BH binaries under the influence of accretion in addition to the scalar/vector and gravitational radiations, and discussed the competition between radiative mechanisms and accretion effects on eccentricity evolution.
However, these discussions were mainly based on the Newtonian orbit and dipole emission.
A generic, fully-relativistic formalism to study EMRIs in spherically symmetric and non-vacuum BH spacetime was established in \cite{Cardoso:2022whc}.
Considering the secondary object of mass $m_p$ orbiting the galactic BHs (GBHs) immersed in an astrophysical environment,
like an accretion disk or a dark matter halo, the authors found that the relative flux difference at $\omega M=0.02$ between a vacuum and a GBH with the halo mass $M_{\text{halo}}=0.1 M$ and the lengthscale $a_0=10^2 M_{\text{halo}}$ and $10^3 M_{\text{halo}}$ is $\sim 10\%$ and $1\%$, respectively.
The results clearly indicate that EMRIs can constrain smaller-scale matter distributions around GBHs \cite{Cardoso:2022whc}.

According to the no-hair theorem, any BH can be described by three parameters: the mass, angular momentum, and electric charge.
Current observations have not yet been able to confirm the  no-hair theorem or the existence of extra fields besides the gravitational fields in modified gravity theories.
The coupling between scalar fields and higher-order curvature invariants can invalidate the no-hair theorem so that BHs can carry scalar charge which depends on the mass of BH \cite{Sotiriou:2013qea,Silva:2017uqg,Doneva:2017bvd,Antoniou:2017acq}.
The possible detection of scalar fields with EMRIs was discussed in \cite{Maselli:2020zgv,Maselli:2021men,Barsanti:2022ana,Guo:2022euk,Zhang:2022rfr,Barsanti:2022vvl,Yunes:2011aa,Cardoso:2011xi}.
%
% Beyond general relativity, BHs can be charged through spontaneous vectorization and the charge of BHs are more conspicuous at larger curvatures \cite{Barton:2021wfj, Oliveira:2020dru, Ramazanoglu:2017xbl}.
% For very different characteristic curvatures near the horizon as one over the mass squared in the EMRI system, the primary object can be taken as chargeless while only the secondary object carries charges.
% So there will be stable vector-charged astrophysical stellar-mass BHs since neutralization arguments for electrical charge do not apply. 
% 
Astrophysical BHs are usually assumed to be neutral because of long-time charge dissipation through the presence of the plasma around them or the spontaneous production of electron-positron pairs \cite{Gibbons:1975kk,Eardley:1975kp,1982PhRvD..25.2509H,1982Prama..18..385J,Gong:2019aqa}. However, the existence of stable charged astrophysical stellar-mass compact objects such as BHs, neutron stars, white dwarfs, etc. in nature remains controversial \cite{Bekenstein:1971ej,de1995relativistic,deFelice:1999qp,Ivanov:2002jy,Majumdar:1947eu,zhang_influence_1982,Anninos:2001yb,Bonnor:1975gba}.
There might exist a maximal huge amount of charge with the charge-to-mass ratio of the order one for highly compact stars, whose radius is on the verge of forming an event horizon \cite{de1995relativistic,deFelice:1999qp}.
The balance between the attractive gravitational force from the matter part and the repulsive force from the electrostatic part is unstable and charged compact stars will collapse to a charged BH due to a decrease in the electric field \cite{Ray:2003gt}.
BHs can also be charged through the Wald mechanism by selectively accreting charge in a magnetic field \cite{Wald:1974np}, or by accreting minicharged dark matter beyond the standard model \cite{Cardoso:2016olt,Bolton:2022hpt}.
The dark photon with mass $\sim 8 \times 10^{-14}$ eVc$^{-2}$ is a minimal and
well-motivated extension of the standard model for reconciling hydrodynamical simulations
with Lyman-$\alpha$ forest absorption line widths at redshift $z\simeq 0.1$ \cite{Bolton:2022hpt,Liu:2022wtq,Liu:2020cds}.

Due to vector radiations, the orbits of EMRIs will be affected,
the merger parameters and the merger rate distribution will be affected too  \cite{Jai-akson:2017ldo,Liu:2022wtq,Liu:2020cds}.
The electromagnetic self-force acting on a charged particle in an equatorial circular orbit of Kerr BH was calculated in \cite{Torres:2020fye}.
It showed the dissipative self-force balances with the sum of the electromagnetic flux radiated to infinity and down the BH horizon,
and prograde orbits can stimulate BH superradiance although the superradiance is not sufficient to support floating orbits even at the innermost stable circular orbit (ISCO) \cite{Torres:2020fye}.
Furthermore, the $U(1)$ charge that is not electromagnetic carried by BHs may bias the parameter estimation of  the chirp mass \cite{Christiansen:2020pnv}.
It was shown that GW150914 is compatible with having charge-to-mass ratio as high as 0.3 \cite{Bozzola:2020mjx}.

In \cite{Zhang:2022hbt}, the energy fluxes for both gravitational and electromagnetic waves induced by a charged particle orbiting around a Schwarzschild BH were studied.
It was demonstrated that the electric charge leaves a significant imprint on the phase of GWs and is observable with space-based GW detectors.
The estimation accuracy of the charges may reach $10^{-5}$ by using the analytic kludge (AK) waveform from the inspiral of a charged stellar-mass compact object
into a charged massive black hole in the weak-field regime \cite{Zi:2022hcc}.
In this paper, based on the Teukolsky formalism for BH perturbations \cite{Teukolsky:1973ha,Press:1973zz,Teukolsky:1974yv},
we numerically calculate the energy fluxes for both tensor and vector perturbations induced by a charged particle moving in an equatorial circular orbit around a Kerr BH, the orbital evolution of EMRIs up to the ISCO,
and investigate the capabilities to detect vector charges carried by the secondary compact object with space-based GW detectors such as LISA, TianQin, and Taiji.
We apply the methods of faithfulness and Fisher information matrix (FIM) to assess the capability of space-based GW detectors to detect the vector charge carried by the secondary compact object.
The paper is organized as follows.
In Sec. \ref{sec2}, we introduce the model with vector charge and the Teukolsky perturbation formalism.
In Sec. \ref{sec3}, we give the source terms as well as procedures for solving the inhomogeneous Teukolsky equations.
Then we numerically calculate the energy fluxes for gravitational and vector fields using the Teukolsky and generalized Sasaki-Nakamura (SN) formalisms in the background of a Kerr BH.
In Sec. \ref{sec4}, we give the numerical results of energy fluxes falling onto the horizon and radiated to infinity for gravitational and vector fields,
then we use the dephasing of GWs to constrain the charge.
In Sec. \ref{sec5}, we calculate the faithfulness between GWs with and without vector charge and perform the FIM to estimate the errors of detecting vector charge with LISA, TianQin, and Taiji.
Sec. \ref{sec6} is devoted to conclusions and discussions.
In this paper, we set $c=G=M=1$.

\section{Einstein-Maxwell field equations}
\label{sec2}
The simplest model including vector charges is Einstein-Maxwell theory, which is modeled via the massless vector field
\begin{equation}
S=\int d^4x\frac{\sqrt{-g}}{16\pi}\left[R-\frac{1}{4}F^{\mu\nu}F_{\mu\nu}-A_{\mu}J^{\mu}\right]-S_{\text{matter}}(g_{\mu\nu},\Phi),
\end{equation}
where $R$ is the Ricci scalar, $A_\mu$ is a massless vector field, $F_{\mu\nu}=\nabla_{\mu}A_{\nu}-\nabla_{\nu}A_{\mu}$ is the field strength and $\Phi$ is the matter field, $J^\mu$ is the electric current density.
Varying the action with respect to the metric tensor and the vector field yields the Einstein-Maxwell field equations
\begin{equation}
G^{\mu\nu}=8\pi T^{\mu\nu}_p+8\pi T^{\mu\nu}_e,
\end{equation}
\begin{equation}
\nabla_\nu F^{\mu\nu}=4\pi J^\mu,
\end{equation}
where $G_{\mu\nu}$ is the Einstein tensor, $T_p^{\mu\nu}$ and $T_e^{\mu\nu}$ are the particle's material stress-energy tensor and vector stress-energy tensor, respectively.
Since the amplitude of the vector stress-energy $T_e^{\mu\nu}$ is quadratic in the vector field, the contribution to the background metric from the vector field is second order.
For an EMRI system $(m_p\ll M)$ composing of a small compact object with mass $m_p$ and  charge-to-mass ratio $q$ orbiting around a Kerr BH with mass $M$ and spin $Ma$, we can ignore the contribution to the background metric from the vector field.
The perturbed Einstein and Maxwell equations for EMRIs are
\begin{equation}
G^{\mu\nu}=8\pi T^{\mu\nu}_p,
\end{equation}
\begin{equation}
\nabla_\nu F^{\mu\nu}=4\pi J^\mu,
\end{equation}
where
\begin{equation}
T^{\mu\nu}_p(x)=m_p\int d\tau~u^{\mu}u^\nu\frac{\delta^{(4)}\left[x-z(\tau)\right]}{\sqrt{-g}},
\end{equation}
\begin{equation}
J^{\mu}(x)=q m_p\int d\tau~ u^{\mu}\frac{\delta^{(4)}\left[x-z(\tau)\right]}{\sqrt{-g}},
\end{equation}
and $u^{\mu}$ is the velocity of the particle.
We use the Newman-Penrose formalism \cite{Newman:1966ub} to study perturbations around a Kerr BH induced by a charged particle with mass $m_p$ and charge $q$.
In Boyer-Lindquist coordinate,  the metric of Kerr BHs is
\begin{equation}\label{SBH}
\begin{split}
ds^2=&(1-2r/\varSigma)dt^2+(4ar\sin(\theta)/\varSigma)dtd\varphi-(\varSigma/\Delta)dr^2-\varSigma d\theta^2\\
&-\sin^2{\theta}(r^2+a^2+2a^2r\sin^2{\theta}/\varSigma)d\varphi^2.
\end{split}\end{equation}
where  $\varSigma=r^2+a^2\cos^2{\theta}$, and $\Delta=r^2-2r+a^2$. When $a=0$, the metric reduces to the Schwarzschild metric.
Based on the metric \eqref{SBH}, we construct the null tetrad,
\begin{equation}
\begin{split}
\begin{split}
l^\mu&=[(r^2+a^2)/\Delta,1,0,a/\Delta],\\
n^\mu&=[r^2+a^2,-\Delta,0,a]/(2\varSigma),\\
m^\mu&=[ia\sin{\theta},0,1,i/\sin{\theta}]/(2^{1/2}(r+ia\cos{\theta})),\\
\bar m^\mu&=[-ia\sin{\theta},0,1,-i/\sin{\theta}]/(2^{1/2}(r-ia\cos{\theta})).
\end{split}
\end{split}
\end{equation}
The propagating vector field is described by the two complex quantities,
\begin{equation}
\phi_0=F_{\mu\nu}l^\mu m^\nu,\qquad \phi_2=F_{\mu\nu}\bar{m}^\mu n^\nu.
\end{equation}
The propagating gravitational field is described by the two complex Newman-Penrose variables
\begin{equation}
\begin{split}
\psi_0&=-C_{\alpha\beta\gamma\delta}l^\alpha m^\beta l^\gamma m^\delta    ,\\
\psi_4&=-C_{\alpha\beta\gamma\delta}n^\alpha \bar{m}^\beta n^\gamma \bar{m}^\delta,
\end{split}
\end{equation}
where $C_{\alpha\beta\gamma\delta}$ is the Weyl tensor.
A single master equation for tensor ($s=-2$) and vector ($s=-1$) perturbations was derived as \cite{Teukolsky:1973ha},

\begin{equation}
\label{TB}
\begin{split}
&\left[\frac{(r^2+a^2)^2}{\Delta}-a^2\sin^2{\theta}\right]\frac{\partial^2\psi}{\partial t^2}+\frac{4ar}{\Delta}\frac{\partial^2\psi}{\partial t\partial\varphi}+\left[\frac{a^2}{\Delta}-\frac{1}{\sin^2{\theta}}\right]\frac{\partial^2\psi}{\partial \varphi^2}\\
&\qquad-\Delta^{-s}\frac{\partial}{\partial r}\left(\Delta^{s+1}\frac{\partial\psi}{\partial r}\right)-\frac{1}{\sin\theta}\frac{\partial}{\partial \theta}\left(\sin\theta\frac{\partial\psi}{\partial \theta}\right)-2s\left[\frac{a(r-1)}{\Delta}+\frac{i\cos\theta}{\sin^2{\theta}}\right]\frac{\partial\psi}{\partial \varphi}\\
&\qquad\qquad\qquad\qquad-2s\left[\frac{(r^2-a^2)}{\Delta}-r-ia\cos\theta\right]\frac{\partial\psi}{\partial t}+(s^2\cot^2\theta-s)\psi=4\pi\varSigma T,
\end{split}
\end{equation}
the explicit field $\psi$ and the corresponding source $T$ are given in  Table \ref{source} \cite{Teukolsky:1973ha}.
 \begin{table}[h]
  \centering
  	\begin{tabular}{|p{0.5cm}<{\centering}|p{3.4cm}<{\centering}|p{3.4cm}<{\centering}|}
		\hline
$s$ & $\psi$ &  $T$\\ \hline
-1   & $(r-i a\cos\theta)^{2}\phi_2$ & $(r-i a\cos\theta)^{2}J_2$  \\ \hline
 -2   & $(r-i a\cos\theta)^{4}\psi_4$ &$2(r-i a\cos\theta)^{4}T_4$  \\ \hline
	\end{tabular}
 \caption{The explicit expressions for the field $\psi$ and the corresponding source $T$.}
    \label{source}
\end{table}
In terms of the eigenfunctions ${_{s}}S_{lm}(\theta)$ \cite{Teukolsky:1973ha,Goldberg:1966uu}, the field $\psi$ can be written as   %
\begin{equation}
\psi=\int d\omega \sum_{l,m}R_{\omega lm}(r)~{_{s}}S_{lm}(\theta)e^{-i\omega t+im\varphi},
\end{equation}
where the radial function $R_{\omega lm}(r)$ satisfies the inhomogeneous Teukolsky equation
\begin{equation}
\label{Teukolsky}
\Delta^{-s}\frac{d}{d r}\left(\Delta^{s+1}\frac{d R_{\omega lm}}{d r}\right)-V_{T}(r)R_{\omega lm}=T_{\omega lm},
\end{equation}
the function
\begin{equation}
V_{T}=-\frac{K^2-2is(r-1)K}{\Delta}-4is\omega r+\lambda_{lm\omega},
\end{equation}
 $K=(r^2+a^2)\omega-am$, $\lambda_{lm\omega}$ is the corresponding eigenvalue which can be computed by the BH Perturbation Toolkit \cite{BHPToolkit}, and the source $T_{\omega lm}(r)$ is
\begin{equation}
T_{\omega lm}(r)=\frac{1}{2\pi}\int dt d\Omega ~4\pi \Sigma T ~{_s}S_{lm}(\theta)e^{i\omega t-im\varphi}.
\end{equation}
For the equatorial circular trajectory at $r_0$ under consideration,
the sources are
\begin{equation}
\begin{split}
T^{\mu\nu}_p(x)&=\frac{m_p}{r_0^2}\frac{u^{\mu}u^{\nu}}{u^t}\delta(r-r_0)\delta(\cos\theta)\delta(\varphi-\hat{\omega} t),\\
J^{\mu}(x)&=q\frac{m_p}{r_0^2}\frac{u^{\mu}}{u^t}\delta(r-r_0)\delta(\cos\theta)\delta(\varphi-\hat{\omega} t),
\end{split}
\end{equation}
where $\hat{\omega}$ is the orbital angular frequency.
Geodesic motion in Kerr spacetime
admits three constants of motion: the specific energy $\hat{E}$, the angular momentum $\hat{L}$, and the Carter constant $\hat{Q}$,
and the geodesic equations are
\begin{eqnarray}
m_p\Sigma \frac{d t}{d \tau} &=&\hat{E} \frac{\varpi^{4}}{\Delta}+a \hat{L}\left(1-\frac{\varpi^{2}}{\Delta}\right)-a^{2} \hat{E}\sin^{2}\theta, \label{timequa}\\
m_p\Sigma \frac{d r}{d \tau} &=&\pm \sqrt{V_{r}\left(r_{0}\right)}, \\
m_p\Sigma \frac{d \theta}{d \tau} &=&\pm \sqrt{V_{\theta}\left(\theta\right)}, \\
m_p\Sigma \frac{d \varphi}{d \tau} &=&a \hat{E}\left(\frac{\varpi^{2}}{\Delta}-1\right)-\frac{a^{2} \hat{L}}{\Delta}+ \hat{L} \csc ^{2} \theta,\label{anglequa}
\end{eqnarray}
where $\varpi\equiv\sqrt{r^2+a^2}$, the radial and polar potentials are
\begin{eqnarray}
&V_{r}(r)& =\left(\hat{E} \varpi^{2}-a \hat{L}\right)^{2}-\Delta\left(r^{2}+\left(\hat{L}-a \hat{E}\right)^{2}+\hat{Q}\right), \\
&V_{\theta}(\theta)& = \hat{Q}-\hat{L}^{2} \cot ^{2} \theta-a^{2}\left(1-\hat{E}^{2}\right) \cos ^{2} \theta.
\end{eqnarray}

In the adiabatic approximation, for a quasi-circular orbit on the equatorial plane,
the coordinates $r$ and $\theta$ are considered as constants,
then Eqs. \eqref{timequa}  and \eqref{anglequa} are the remaining equations.
The conserved constants are \cite{Detweiler:1978ge}
\begin{eqnarray}\label{orbitE}
	\hat{E}&=& m_p\frac{r_0^{3 / 2}-2 r_{0}^{1 / 2} \pm a }{r_{0}^{3 / 4}\left(r_{0}^{3 / 2}-3  r_{0}^{1 / 2} \pm 2 a \right)^{1 / 2}},    \\
	\hat{L}&=&m_p \frac{\pm (r_{0}^{2}\mp 2a r_{0}^{1 / 2} + a^2)}{r_{0}^{3 / 4}\left(r_{0}^{3 / 2}-3  r_{0}^{1 / 2} \pm 2 a \right)^{1 / 2}},\\
	\hat{Q}&=&0.
\end{eqnarray}
The orbital angular frequency is
\begin{equation}\label{orbitF}
\hat{\omega} \equiv \frac{d\varphi}{dt}=\frac{\pm 1}{r_{0}^{3 / 2} \pm a},
\end{equation}
where $\pm$ corresponds to  co-rotating and counter-rotating, respectively.
In the following discussions, we use positive $a$ for co-rotating cases and negative $a$ for counter-rotating cases.

\section{Numerical calculation for the energy flux}
\label{sec3}
The homogeneous Teukolsky equation \eqref{Teukolsky} admits two linearly independent solutions $R^{\text{in}}_{\omega lm}$ and $R^{\text{up}}_{\omega lm}$, with the following asymptotic values at the horizon $r_+$ and at infinity,
\begin{equation}
R^{\text{in}}_{\omega lm}=
\begin{cases}
B^{\text{tran}}\Delta^{-s}e^{-i\kappa r^*},&\quad (r\to r_+)\\
B^{\text{out}}\frac{e^{i\omega r^*}}{r^{2s+1}}+B^{\text{in}}\frac{e^{-i\omega r^*}}{r}, &\quad (r\to+\infty)
\end{cases}
\end{equation}
\begin{equation}
R^{\text{up}}_{\omega lm}=
\begin{cases}
D^{\text{out}}e^{i\kappa r^*}+\frac{D^{\text{in}}}{\Delta^{s}}e^{-i\kappa r^*},&\quad (r\to r_+)\\
D^{\text{tran}}\frac{e^{i\omega r^*}}{r^{2s+1}},&\quad (r\to+\infty)\\
\end{cases}
\end{equation}
where $\kappa=\omega-m a/(2r_+)$, $r_\pm=1\pm\sqrt{1-a^2}$, and the tortoise radius of the Kerr metric
\begin{equation}
r^*=r+\frac{2r_+}{r_+-r_-}\ln \frac{r-r_+}{2}-\frac{2r_-}{r_+-r_-}\ln \frac{r-r_-}{2}.
\end{equation}
The solutions $R^{\text{in}}_{\omega lm}$ and $R^{\text{up}}_{\omega lm}$  are purely outgoing at infinity and purely ingoing at the horizon.
With the help of these homogeneous solutions, the solution to Eq.~\eqref{Teukolsky} is

\begin{equation}
\begin{split}
R_{\omega lm}(r)=\frac{1}{W}
\left(R^{\text{in}}_{\omega lm}\int_{r}^{+\infty}\Delta^{s}R^{\text{up}}_{\omega lm}T_{\omega lm}dr+R^{\text{up}}_{\omega lm}\int_{r_+}^{r}\Delta^{s}R^{\text{in}}_{\omega lm}T_{\omega lm}dr\right).
\end{split}
\end{equation}
with the constant Wronskian given by
\begin{equation}
W= \Delta^{s+1} \left(R^{\text{in}}_{\omega lm}\frac{d R^{\text{up}}_{\omega lm}}{dr}-R^{\text{up}}_{\omega lm}\frac{d R^{\text{in}}_{\omega lm}}{dr}\right)=2i\omega B^{\text{in}} D^{\text{tran}}.
\end{equation}
The solution is purely outgoing at infinity and purely ingoing at the horizon,
\begin{equation}
\begin{split}
R_{\omega lm}(r\to r_+)=Z^{\infty}_{\omega lm}\Delta^{-s}e^{-i\kappa r^*},\\
R_{\omega lm}(r\to \infty)=Z^{H}_{\omega lm}r^{-2s-1}e^{i\omega r^*},
\end{split}
\end{equation}
with
\begin{equation}
\begin{split}
Z^{\infty}_{\omega lm}&=\frac{B^{\text{tran}}}{W}\int_{r_+}^{+\infty}\Delta^{s}R^{\text{up}}_{\omega lm}T_{\omega lm}dr,\\
Z^{H}_{\omega lm}&=\frac{D^{\text{tran}}}{W}\int_{r_+}^{+\infty}\Delta^{s}R^{\text{in}}_{\omega lm}T_{\omega lm}dr.
\label{amplitudes}
\end{split}
\end{equation}
For a circular equatorial orbit with orbital angular frequency $\hat{\omega} $, we get
\begin{equation}
Z^{H,\infty}_{\omega lm}=\delta(\omega-m \hat{\omega})\mathcal{A}^{H,\infty}_{\omega lm}.
\end{equation}
For $s=-1$, the energy fluxes at infinity and the horizon read
\begin{equation}
\begin{split}
\dot{E}_q^{\infty}=\left(\frac{d E}{dt}\right)_{EM}^\infty&=\sum_{l=1}^{\infty}\sum_{m=1}^{l}\frac{|\mathcal{A}^{H}_{\omega lm}|^2}{\pi}, \\
\dot{E}_q^H=\left(\frac{d E}{dt}\right)_{EM}^H&=\sum_{l=1}^{\infty}\sum_{m=1}^{l}\alpha^E_{lm}\frac{|\mathcal{A}^{\infty}_{\omega lm}|^2}{\pi},
\end{split}
\end{equation}
where the coefficient $\alpha^E_{lm}$ is \cite{Teukolsky:1974yv}
\begin{equation}\label{energyformula}
\alpha^E_{lm}=\frac{128\omega\kappa r_+^3(\kappa^2+4\epsilon^2)}{|B_E|^2}
\end{equation}
with $\epsilon=\sqrt{1-a^2}/(4r_+)$ and
\begin{equation}
|B_E|^2=\lambda_{lm\omega}^2+4ma\omega-4a^2\omega^2.
\end{equation}
For $s=-2$, the gravitational energy fluxes at infinity and the horizon are given by
\begin{equation}
\begin{split}
\dot{E}_{\text{grav}}^{\infty}=\left(\frac{d E}{dt}\right)_{GW}^\infty&=\sum_{l=2}^{\infty}\sum_{m=1}^{l}\frac{|\mathcal{A}^{H}_{\omega lm}|^2}{2\pi\omega^2}, \\
\dot{E}_{\text{grav}}^H=\left(\frac{d E}{dt}\right)_{GW}^H&=\sum_{l=2}^{\infty}\sum_{m=1}^{l}\alpha^G_{lm}\frac{|\mathcal{A}^{\infty}_{\omega lm}|^2}{2\pi\omega^2},
\end{split}
\end{equation}
where the coefficient $\alpha^G_{l m}$ is \cite{Hughes:1999bq}
\begin{equation}\label{energyformula2}
\alpha^G_{l m}=\frac{256\left(2 r_{+}\right)^5 \kappa\left(\kappa^2+4 \epsilon^2\right)\left(\kappa^2+16 \epsilon^2\right)\omega^3}{\left|B_G\right|^2},
\end{equation}
and
\begin{equation}
\begin{aligned}
\left|B_G\right|^2 &=\left[\left(\lambda_{l m \omega}+2\right)^2+4 a\omega-4 a^2\omega^2\right]\times\left[\lambda_{l m \omega}^2+36 m a\omega-36 a^2\omega^2\right] \\
&+\left(2 \lambda_{l m \omega}+3\right)\left[96 a^2\omega^2-48 m a\omega\right]+144\omega^2\left(1-a^2\right) .
\end{aligned}
\end{equation}
Therefore, the total energy fluxes emitted from the EMRIs read
\begin{equation}
\dot{E}=    \dot{E}_q+\dot{E}_{\text{grav}},
\end{equation}
where
\begin{equation}
    \dot{E}_q=\dot{E}_q^\infty+\dot{E}_q^H,\ \ \dot{E}_{\text{grav}}=\dot{E}_{\text{grav}}^\infty+\dot{E}_{\text{grav}}^H.
\end{equation}
The detailed derivation of the above results is given in Appendix \ref{gsne}.
The energy flux emitted by tensor fields can be computed with the BH Perturbation Toolkit \cite{BHPToolkit}.

\section{Results}\label{sec4}
\begin{figure}
    \centering
    \includegraphics[width=0.95\columnwidth]{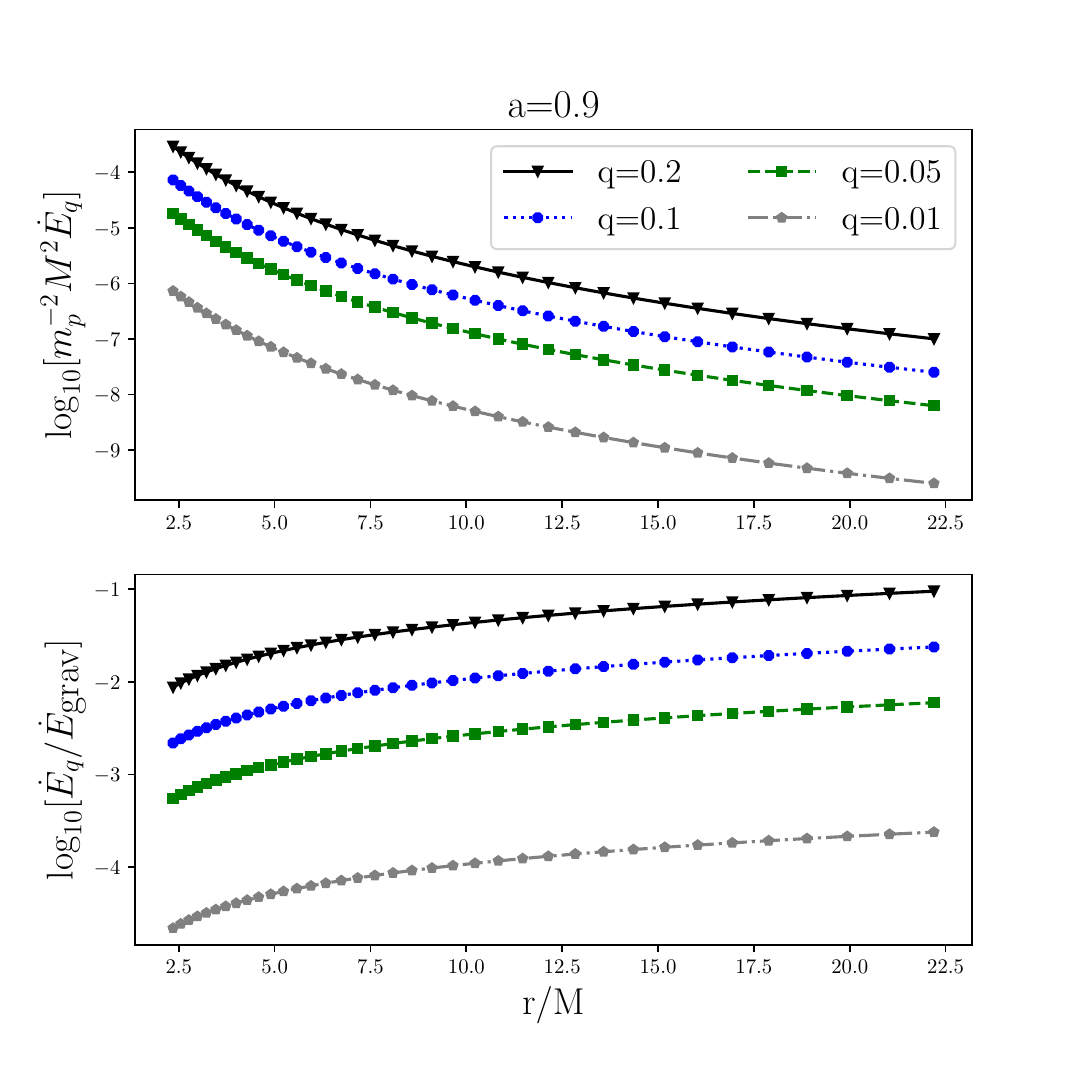}
    \caption{The energy fluxes versus the orbital distance.
    The top panel shows the vector flux normalized with the mass ratio from a charged particle orbiting around a Kerr BH with the spin of $a/M=0.9$ for different values of the vector charge $q$.
    The bottom panel shows the ratio between vector and gravitational energy fluxes for different values of the vector charge $q$.}
    \label{energyd}
\end{figure}

\begin{figure}
    \centering
    \includegraphics[width=0.95\columnwidth]{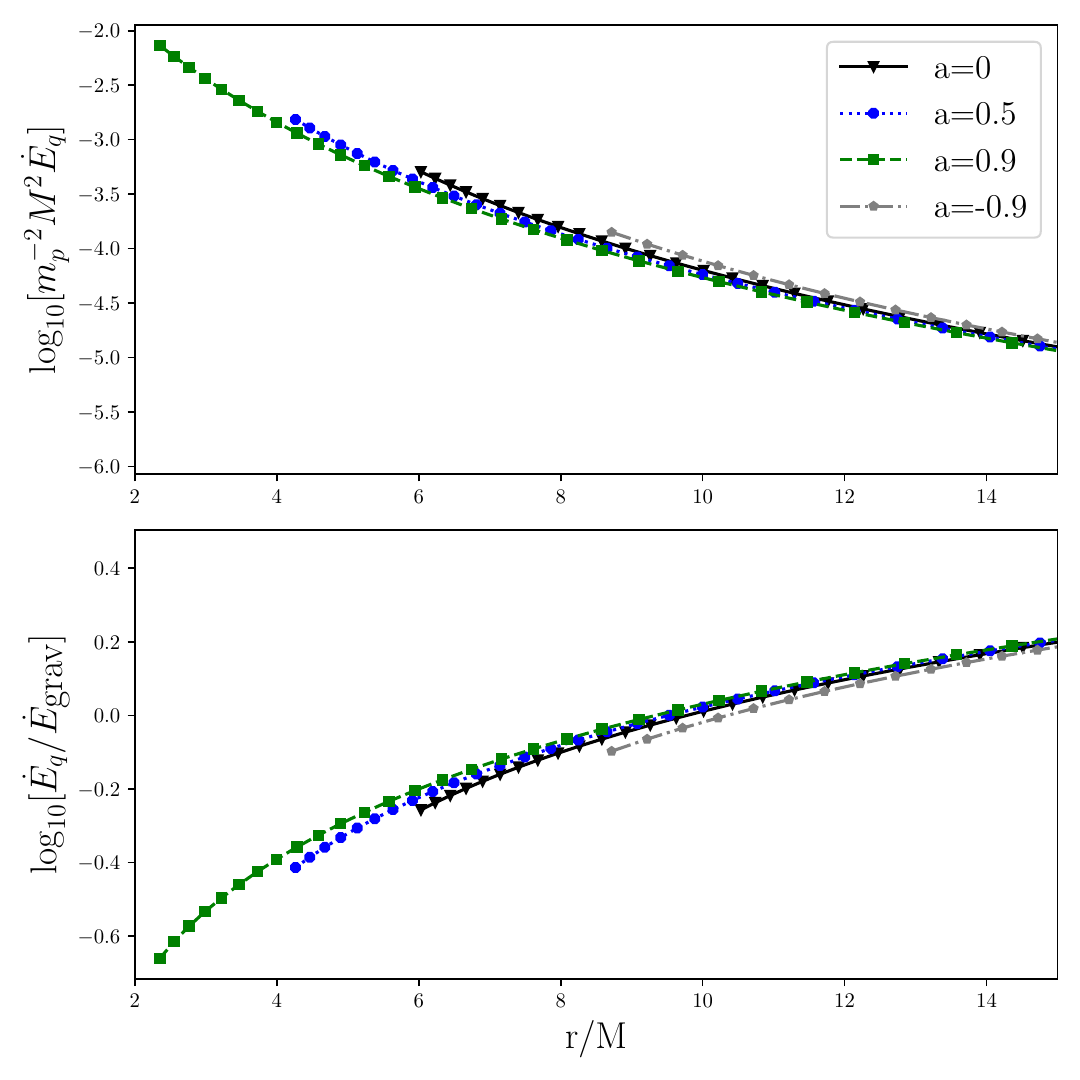}
    \caption{Same as Fig. \ref{energyd}, but for different values of the primary spin $a$.
    The vector charge $q$ is set to 1.}
    \label{energya}
\end{figure}

\begin{figure}
    \centering
    \includegraphics[width=0.95\columnwidth]{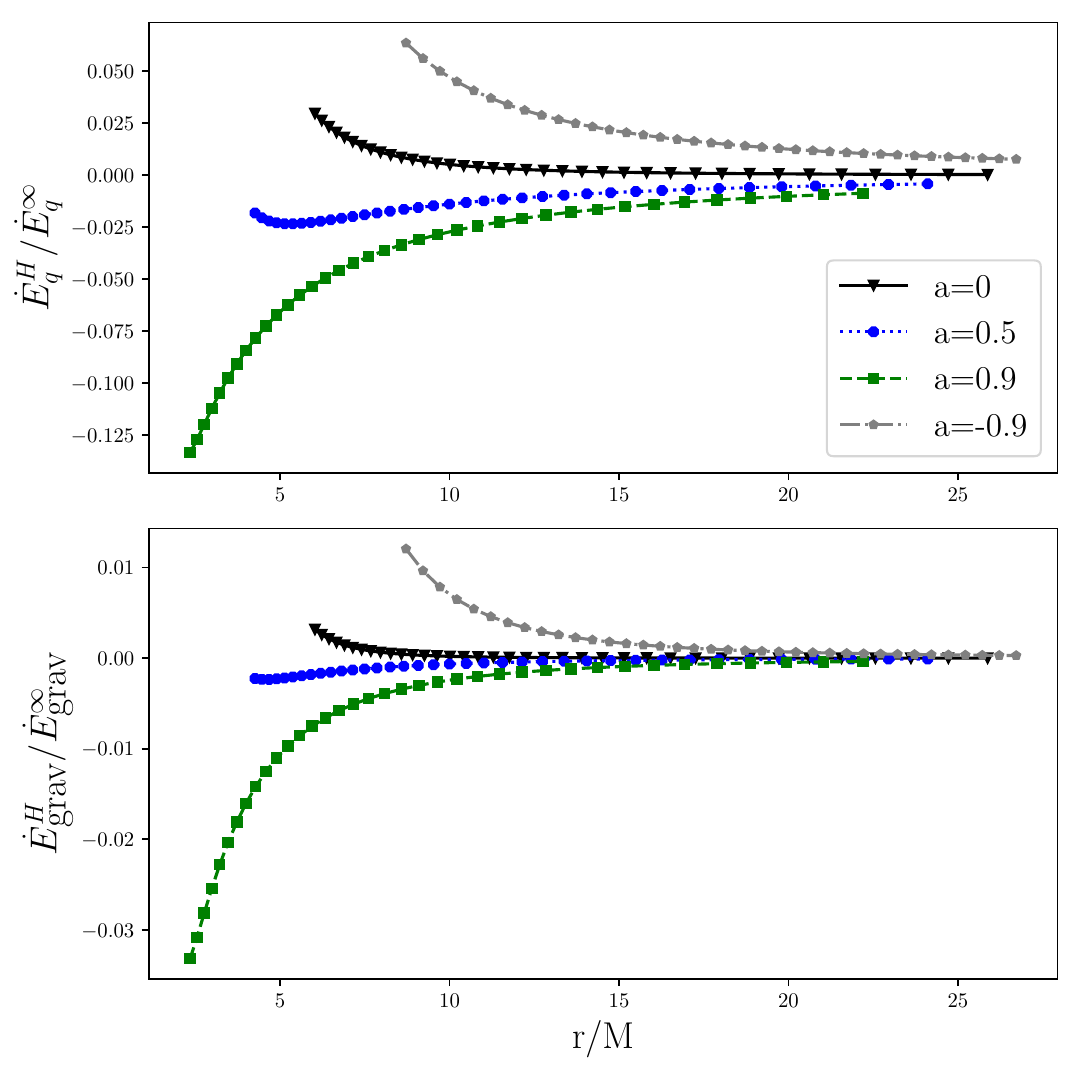}
    \caption{The ratio of the energy flux falling onto the horizon to the energy flux radiated to infinity, as a function of the orbital radius, for various spin $a$, and for the vector (the top panel) and gravitational cases (the bottom panel).}
    \label{energyHI}
\end{figure}

The top  panel of Fig. \ref{energyd} shows the normalized vector energy flux $m_p^{-2}M^2\dot{E}_q$ for a charged particle with different charge values of $q$ on a circular orbit about a Kerr BH with the spin $a=0.9$, as a function of orbital radius.
The vector energy flux is proportional to the square of the vector charge $q^2$.
The vector energy flux increases as the charged particle inspirals into the central Kerr BH.
The ratio between the vector and gravitational energy flux is shown in the bottom panel of Fig. \ref{energyd}.
Both the vector and gravitational fluxes are in the same order of $(m_p/M)^2$,
the ratio of fluxes is independent of the mass ratio and increases as the orbital radius because the gravitational contribution falls off faster than the vector energy flux at a large orbital radius.
Figure \ref{energya} shows the normalized vector energy flux $m_p^{-2}M^2\dot{E}_q$ and the ratio of energy fluxes $\dot{E}_q/\dot{E}_{\textrm{grav}}$ as a function of the orbital radius for different values of $a$.
For larger $a$, the ISCO is smaller, resulting in higher GW frequency at the coalescence.
For the same orbital radius, the vector energy flux is slightly larger for smaller $a$.
However, the total energy flux increases with $a$ for one-year observations before the merger due to the smaller ISCO.
Figure \ref{energyHI} shows the ratio of energy flux falling onto the horizon to the energy flux radiated away to infinity,
as a function of the orbital radius, for various spin  $a$, and for the vector and gravitational fields.
It is interesting to note that the sign of ratio becomes negative for Kerr BH with positive $a$ (co-rotating orbit) at  a small orbital radius.
In these cases, the vector and gravitational fields generate superradiance, leading to extraction of energy from the horizon.
The superradiance only happens when the coefficient $\kappa$ in Eqs. \eqref{energyformula} and \eqref{energyformula2} becomes negative, which means that the orbital frequency slows down the rotation of the Kerr BH.
Our results are consistent with those found in Ref. \cite{Torres:2020fye}.

\begin{figure}
    \centering
    \includegraphics[width=0.95\columnwidth]{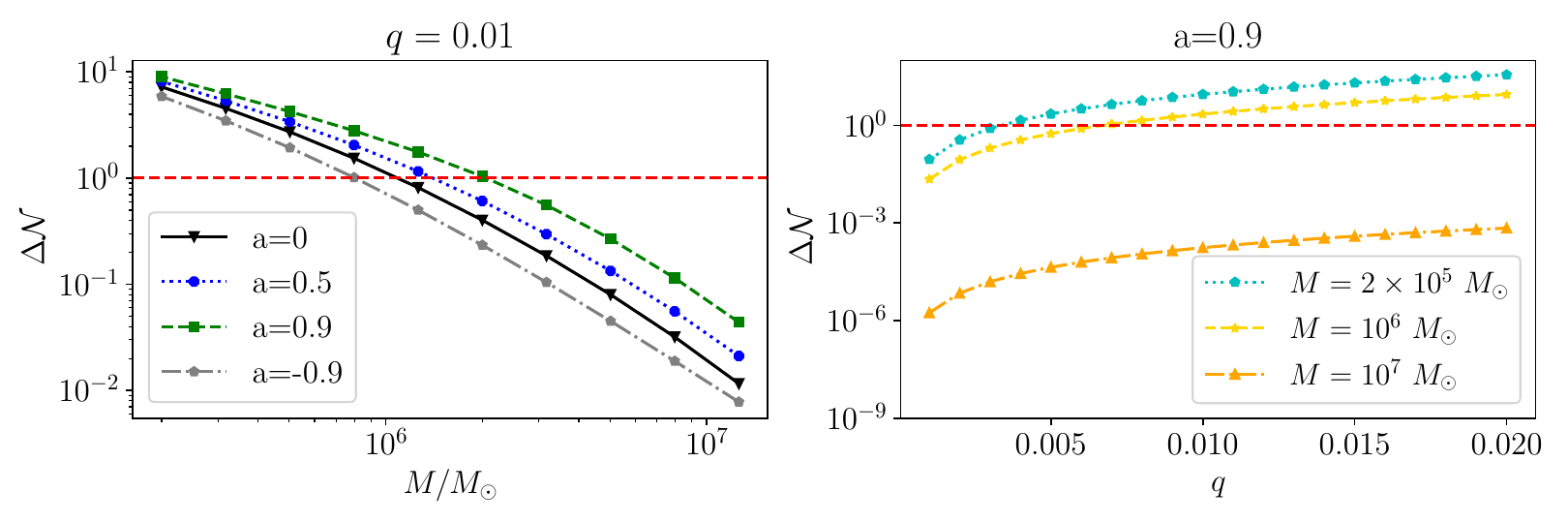}
    \caption{The difference between the number of GW cycles accumulated by EMRIs with and without the vector charge in circular orbits. The left panel shows the dephasing as a function of the mass of the Kerr BH in the range $M\in\left[2\times 10^5,2\times 10^7\right]M_\odot$ for different spin values, the charge $q=0.01$.
    The right panel shows the dephasing as a function of the charge-to-mass ratio $q$ for different $M$, the spin $a=0.9$. The red dashed line corresponds to the threshold above which two signals are distinguishable with space-based GW detectors. All observational time is one year before the merger.}
    \label{phase}
\end{figure}

The extra energy leakage due to the vector field accelerates the coalescence of binaries.
Therefore, we expect the vector charge to leave a significant imprint on the GW phase over one-year evolution before the merger for EMRIs.
To detect the vector charge carried by the small compact object in EMRIs,
we study the dephasing of GWs caused by the additional energy loss during inspirals.
The observation time is one year before the merger,
\begin{equation}\label{sol-fre}
	T_{\text{obs}}=\int^{f_{\text{max}}}_{f_{\text{min}}}\frac{1}{\dot{f}}df=1\ \text{year},
\end{equation}
where
\begin{equation}
    f_{\text{max}}=\text{min}(f_{\text{ISCO}},f_{\text{up}}),~~~~~~f_{\text{min}}=\text{max}(f_{\text{low}},f_{\text{start}}),
\end{equation}
$f=\hat{\omega}/\pi$ is the GW frequency, $f_{\text{ISCO}}$ is the frequency at the ISCO \cite{Jefremov:2015gza},
$f_{\text{start}}$ is the initial frequency at $t=0$,
the cutoff frequencies $f_{\text{low}}=10^{-4}$ Hz and $f_{\text{up}}=1$ Hz.
The orbit evolution is determined by
 \begin{equation}\label{orbittime}
 \frac{d r}{dt}=-\dot{E}\left(\frac{d \hat{E}}{dr}\right)^{-1},\qquad \frac{d \varphi_{\text{orb}}}{d t}=\pi f,
 \end{equation}
 where $\dot{E}=\dot{E}_q+\dot{E}_{\text{grav}}$.
The total number of GW cycles accumulated over one year before the merger is \cite{Berti:2004bd}
\begin{equation}\label{phase-end}
\mathcal{N}=\int_{f_{\text{min}}}^{f_{\max }} \frac{f}{\dot{f}} d f.
\end{equation}
Considering EMRIs with the mass of the second compact object being fixed to be $m_p=10~M_{\odot}$,
we calculate the dephasing $\Delta\mathcal{N}=\mathcal{N}(q=0)-\mathcal{N}(q)$ for different vector charge $q$, spin $a$ and mass $M$,
and the results are shown in Fig. \ref{phase}.
% As discussed above, the extra emission channel due to extra electromagnetic radiation accelerates the coalescence,
For one-year observations before the merger,
the charged particle starts further away from ISCO due to extra radiation of the vector field and the difference $\Delta\mathcal{N}$ is always positive.
As shown in Fig. \ref{phase}, $\Delta \mathcal{N}$ increases monotonically with the spin $a$ and the charge-to-mass ratio $q$, and it strongly depends on the mass of the central BH such that lighter BHs have larger $\Delta \mathcal{N}$.
This means that the observations of EMRIs with a lighter and larger-spin Kerr BH can detect the vector charge easier.
For the same EMRI configuration in the Kerr background, the co-rotating orbit can detect the vector charge easier than the counter-rotating orbit.
Following Refs. \cite{Berti:2004bd,Maselli:2020zgv},
we take the threshold for a detectable dephasing that two signals are distinguishable by space-based GW detectors as $\Delta \mathcal{N}=1$.
Observations of EMRIs over one year before the merger may be able to reveal the presence of a vector charge as small as $q\sim 0.007$ for Kerr BHs with $a=0.9$ and $M=10^6~M_\odot$, and $q\sim 0.01$ for Schwarzschild BHs with $M=10^6~M_\odot$.

\section{Parameter Estimation}
\label{sec5}
To make the analysis more accurate and account for the degeneracy among parameters,
we calculate the faithfulness between two GW waveforms and carry out parameter estimation with the FIM method.
\subsection{Signals}
\begin{figure}
	\centering
	\includegraphics[width=0.8\columnwidth]{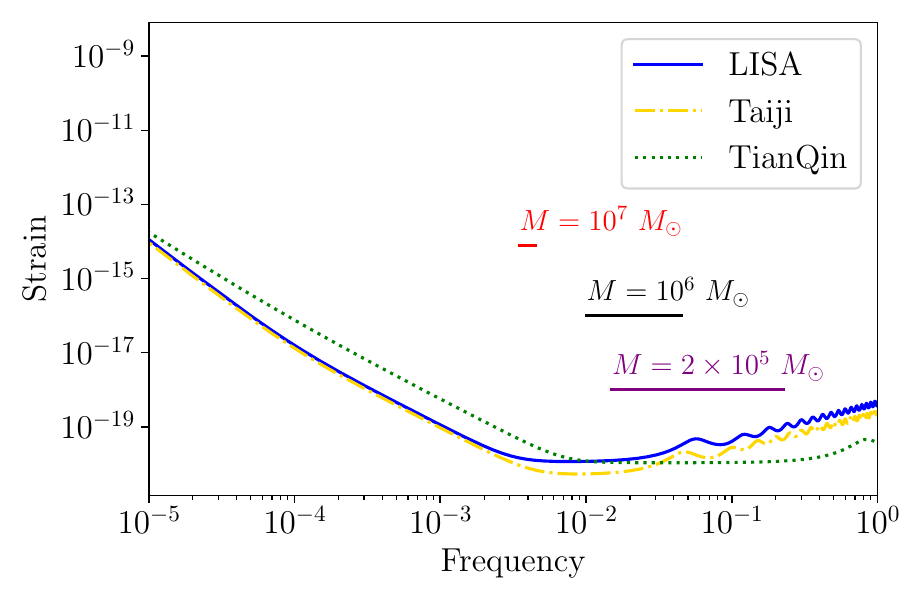}
	\caption{The sensitivity curves for LISA, TianQin, and Taiji.
The horizontal solid lines represent the frequency band $f_{\textrm{start}}$ to $f_{\textrm{ISCO}}$ for EMRIs with $a=0.9$, and $M=2\times10^5~M_{\odot}$, $M=10^6~M_{\odot}$ and $M=10^7~M_{\odot}$ over one-year evolution before the merger.
}
\label{sensitivity}
\end{figure}

\begin{figure}[thbp]
\center{
\includegraphics[scale=0.74]{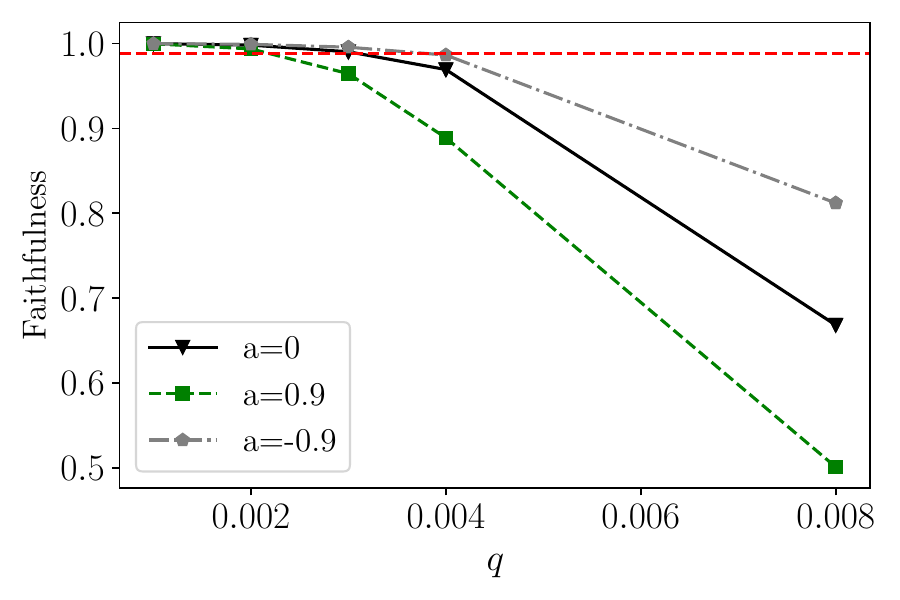}
\caption{Faithfulness between GW signals with and without the vector charge for LISA as a function of the charge $q$. The spin of the Kerr BH is $a=0$, $a=0.9$, and $a=-0.9$. The horizontal dashed line represents the detection limit with LISA, $\mathcal{F}_n= 0.988$.
}\label{faithfulness}}
\end{figure}

We can obtain the inspiral trajectory from adiabatic evolution in Eq. \eqref{orbittime}, then compute GWs in the quadrupole approximation.
The metric perturbation in the transverse-traceless (TT) gauge is
\begin{equation}
h_{i j}^{\mathrm{TT}}=\frac{2}{d_L}\left(P_{i l} P_{j m}-\frac{1}{2} P_{i j} P_{l m}\right) \ddot{I}_{l m},
\end{equation}
where $d_L$ is the luminosity distance of the source,
$P_{ij}=\delta_{ij}-n_i n_j$ is the projection operator acting onto GWs with the unit propagating direction $n_j$,
$\delta_{ij}$ is the Kronecker delta function,
and $\ddot{I}_{ij}$ is the second time derivative of the mass quadrupole moment.
The GW strain measured by the detector is
\begin{equation}\label{signal}
h(t)=h_{+}(t) F^{+}(t)+h_{\times}(t) F^{\times}(t),
\end{equation}
where $h_+(t)=\mathcal{A}\cos\left[2\varphi_{\rm orb}+2\varphi_0\right]\left(1+\cos^2\iota\right)$, $h_\times(t)=-2\mathcal{A}\sin\left[2\varphi_{\rm orb}+2\varphi_0\right]\cos\iota$,  $\iota$ is the inclination angle between the binary orbital angular momentum and the line of sight,
the GW amplitude  $\mathcal{A}=2m_{\rm p}\left[M\hat{\omega}(t)\right]^{2/3}/d_L$ and $\varphi_0$ is the initial phase.
The interferometer pattern functions $F^{+,\times}(t)$ and $\iota$ can be expressed in terms of four angles which specify the source orientation, $(\theta_s,\phi_s)$,
and the orbital angular direction $(\theta_1,\phi_1)$.
% Their explicit expression is given in Appendix \ref{detectors}.
The faithfulness between two signals is defined as
\begin{equation}\label{eq:def_F}
\mathcal{F}_n[h_1,h_2]=\max_{\{t_c,\phi_c\}}\frac{\langle h_1\vert
	h_2\rangle}{\sqrt{\langle h_1\vert h_1\rangle\langle h_2\vert h_2\rangle}}\ ,
\end{equation}
where $(t_c,\phi_c)$ are time and phase offsets \cite{Lindblom:2008cm},
the noise-weighted inner product between two templates $h_1$ and $h_2$ is
\begin{equation}\label{product}
\left\langle h_{1} \mid h_{2}\right\rangle=4 \Re \int_{f_{\min }}^{f_{\max }} \frac{\tilde{h}_{1}(f) \tilde{h}_{2}^{*}(f)}{S_{n}(f)} df,
\end{equation}
$\tilde{h}_{1}(f)$ is the Fourier transform of the time-domain signal $h(t)$,
its complex conjugate is $\tilde{h}_{1}^{*}(f)$,
and $S_n(f)$ is the noise spectral density  for space-based GW detectors.
The signal-to-noise ratio (SNR) can be obtained by calculating $\rho=\left\langle h|h \right\rangle^{1/2}$.
The sensitivity curves of LISA, TianQin, and Taiji are shown in Fig. \ref{sensitivity}.
As pointed out in \cite{Chatziioannou:2017tdw},
two signals can be distinguished by LISA if $\mathcal{F}_n\leq0.988$.
Here we choose the source masses $m_p=10~M_{\odot}$, $M=10^6~M_{\odot}$,
the source angles $\theta_s=\pi/3,~\phi_s=\pi/2$ and $\theta_1=\phi_1=\pi/4$,
the luminosity distance is scaled to ensure SNR $\rho=30$,
the initial phase is set as $\varphi_0=0$ and the initial orbital separation is adjusted to experience one-year adiabatic evolution before the plunge $r_{\text{end}}=r_{\text{ISCO}}+0.1~M$.
In Fig. \ref{faithfulness}, we show
the faithfulness between GW signals with and without the vector charge for LISA as a function of the vector charge.
The results show that one-year observations of EMRIs with LISA may be able to reveal the presence of a vector charge as small as $q\sim 0.002$ for Kerr BHs with $a=0.9$ and $M=10^6~M_\odot$ (co-rotating orbit),
$q\sim 0.003$ for Schwarzschild BHs with $a=0$ and $M=10^6~M_\odot$, and $q\sim 0.004$ for Kerr BHs with $a=-0.9$ and $M=10^6~M_\odot$ (counter-rotating orbit).
Larger positive spin of the Kerr BH (co-rotating orbit) can help us detect the vector charge easier, which is consistent with the results obtained from the dephasing in the previous section.

\subsection{Fisher information matrix}
The signals \eqref{signal} measured by the detector are determined by the following eleven parameters
\begin{equation}
\xi=(\ln M, \ln m_p, a, q, r_0, \varphi_0, \theta_s, \phi_s, \theta_1, \phi_1, d_L).
\end{equation}
 In the large SNR limit,
the posterior probability distribution of the source parameters $\xi$ can be approximated by a multivariate Gaussian distribution centered around the true values $\hat{\xi}$.
Assuming flat or Gaussian priors on the source parameters $\xi$,
their covariances are given by the inverse of the FIM
\begin{equation}
\Gamma_{i j}=\left\langle\left.\frac{\partial h}{\partial \xi_{i}}\right| \frac{\partial h}{\partial \xi_{j}}\right\rangle_{\xi=\hat{\xi}}.
\end{equation}
The statistical error on $\xi$ and the correlation coefficients between the parameters are provided by the diagonal and non-diagonal parts of ${\bf \Sigma}={\bf \Gamma}^{-1}$, i.e.
\begin{equation}
\sigma_{i}=\Sigma_{i i}^{1 / 2} \quad, \quad c_{\xi_{i} \xi_{j}}=\Sigma_{i j} /\left(\sigma_{\xi_{i}} \sigma_{\xi_{j}}\right).
\end{equation}
Because of the triangle configuration of the space-based GW detector, the total SNR is defined by $\rho=\sqrt{\rho_1^2+\rho_2^2}$, so the total covariance matrix of the binary parameters is obtained by inverting the sum of the Fisher matrices $\sigma_{\xi_i}^2=(\Gamma_1+\Gamma_2)^{-1}_{ii}$.
Here we fix the source angles $\theta_s=\pi/3,~\phi_s=\pi/2$ and $\theta_1=\phi_1=\pi/4$,
the initial phase is set as $\varphi_0=0$ and the initial orbital separation is adjusted to experience one-year adiabatic evolution before the plunge $r_{\text{end}}=r_{\text{ISCO}}+0.1~M$.
The luminosity distance $d_L$ is set to be $1$ Gpc.
We apply the FIM method for LISA, TianQin, and Taiji to estimate the errors of the vector charge.

\begin{figure}
\centering
\includegraphics[width=0.6\columnwidth]{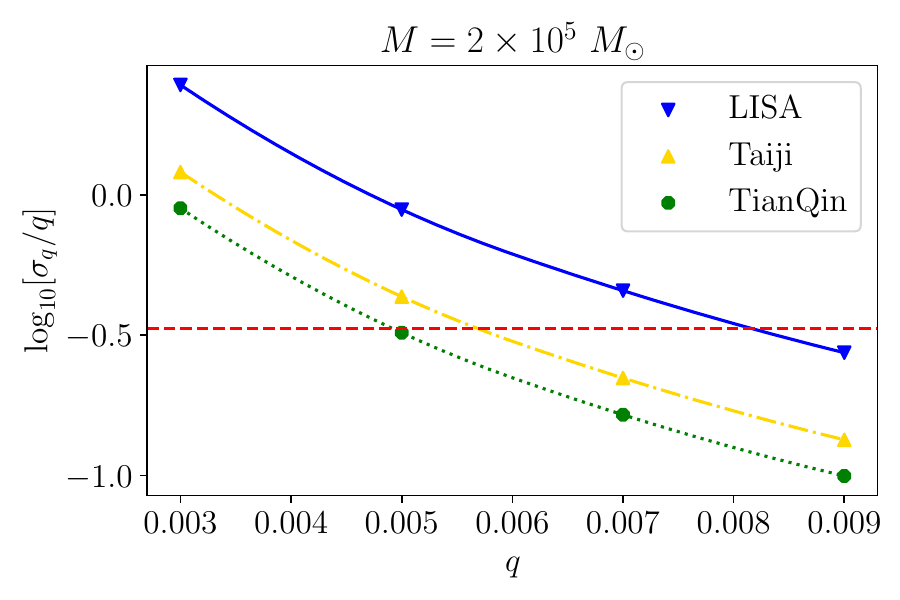}\\
\includegraphics[width=0.6\columnwidth]{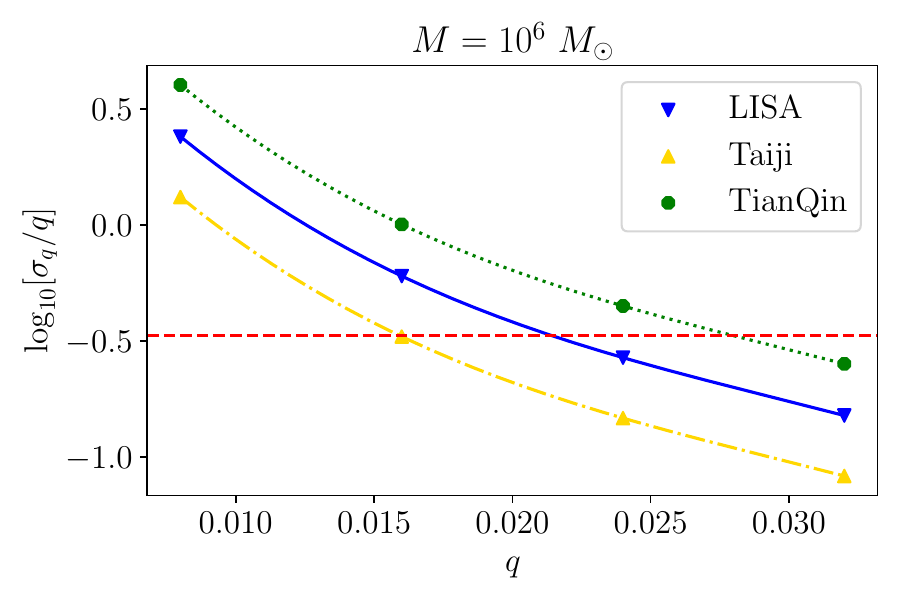}\\
\includegraphics[width=0.6\columnwidth]{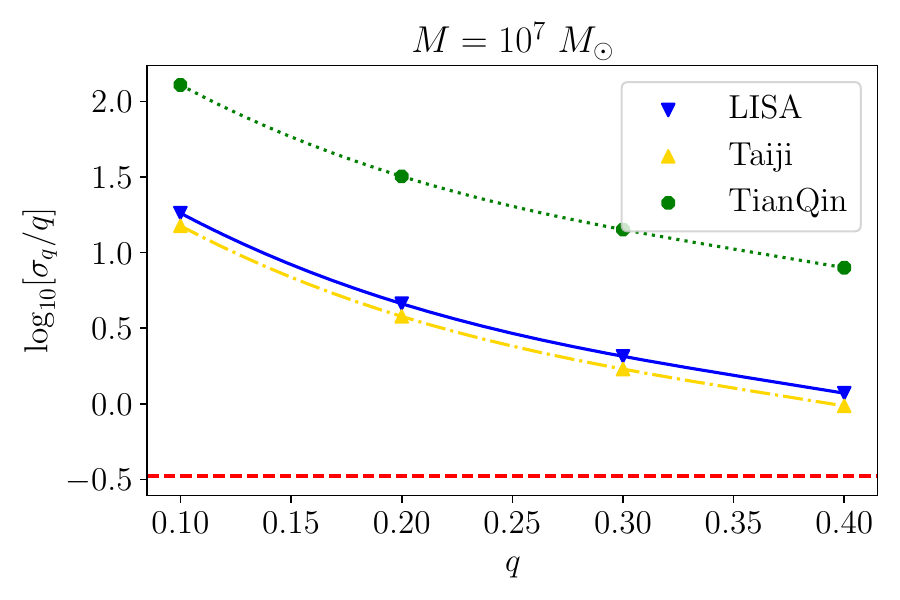}\\
\caption{
The $1\sigma$ interval for the charge $q$ as a function of the charge $q$, inferred after one-year observations of EMRI with $a=0.9$ and different $M$ with LISA, Taiji, and TianQin. The horizontal dashed lines represent the $3\sigma$ limit $33.3\%$.
}
\label{sigmaq}
\end{figure}

The relative errors of the vector charge $q$ as a function of the vector charge with LISA, TianQin and Taiji are shown in Fig. \ref{sigmaq}.
For one-year observations before the merger,
the charged particle starts further away from ISCO due to extra radiation of the vector field,
so the $1\sigma$ error for the charge decreases with the charge $q$.
For EMRIs with $M=2\times 10^5~M_{\odot}$ and $a=0.9$, as shown in the top panel,
the relative errors of the charge $q$ with TianQin are better than LISA and Taiji.
For $M=10^6~M_{\odot}$ and $a=0.9$ as shown in the middle panel,
the relative errors of the charge $q$ with TianQin and LISA are almost the same.
For $M=10^7~M_{\odot}$ and $a=0.9$ as shown in the bottom panel,
the relative errors of the charge $q$ with TianQin are worse than LISA and Taiji.
In all the cases, the relative errors of the charge $q$ with Taiji are better than LISA, for the reason that the sensitivity of Taiji is always better than LISA.
For $M=2\times 10^5~M_{\odot}$ and $a=0.9$,
the relative errors with TianQin are better than LISA and Taiji since the sensitivity of TianQin is better than LISA and Taiji in the high-frequency band, but worse than LISA and Taiji in the low-frequency band as shown in Fig. \ref{sensitivity}.
For EMRIs with $m_p=10~M_{\odot}$, $M=10^6~M_{\odot}$ and $a=0.9$,
the vector charge can be constrained for LISA as small as $q\sim0.021$,
for TianQin as small as $q\sim0.028$ and for Taiji as small as $q\sim0.016$.
For EMRIs with $m_p=10~M_{\odot}$, $M=2\times10^5~M_{\odot}$ and $a=0.9$,
the vector charge can be constrained for LISA as small as $q\sim0.0082$,
for TianQin as small as $q\sim0.0049$ and for Taiji as small as $q\sim0.0057$.

\begin{figure}
\centering
\includegraphics[width=0.6\columnwidth]{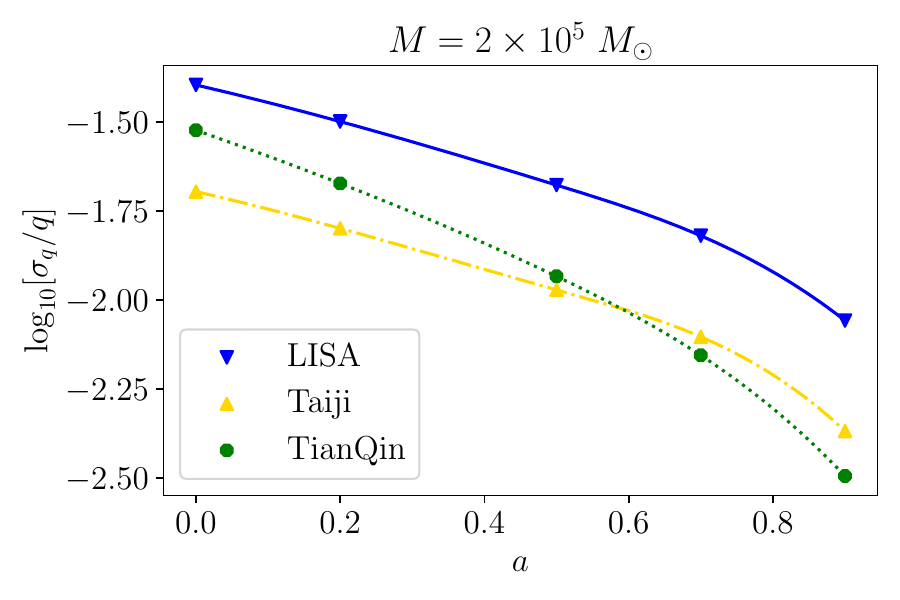}\\
\includegraphics[width=0.6\columnwidth]{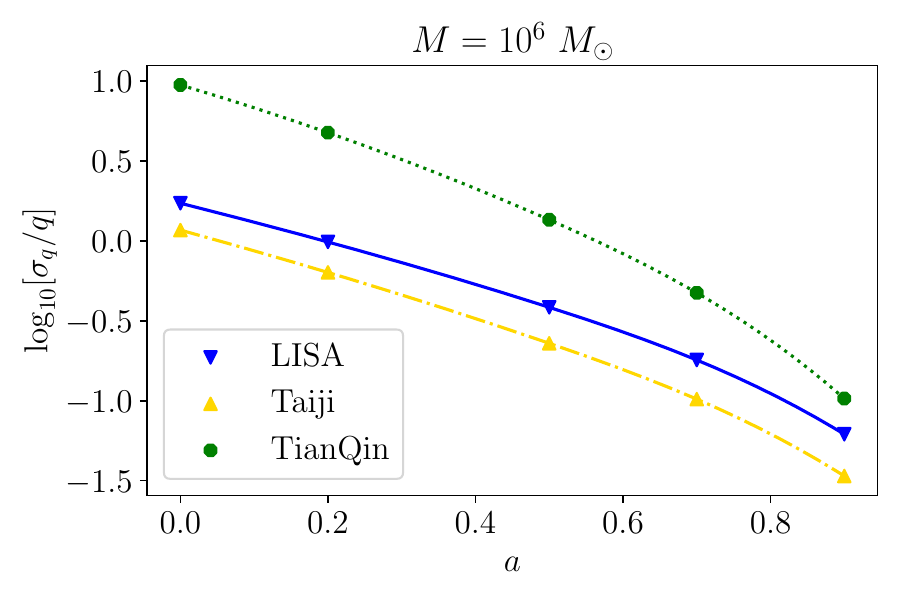}\\
\includegraphics[width=0.58\columnwidth]{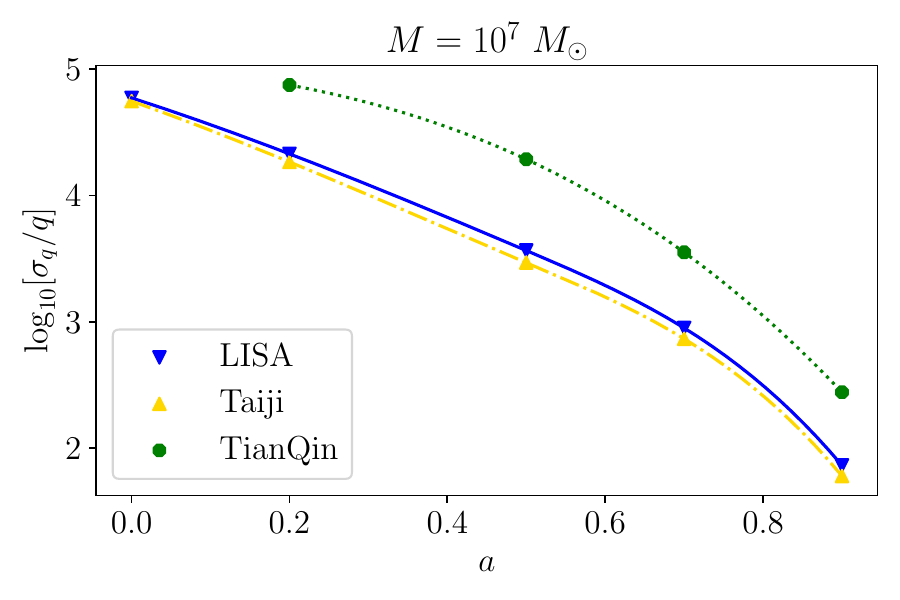}\\
\caption{
The $1\sigma$ interval for the charge $q$ as a function of the spin $a$, inferred after one-year observations of EMRI with $q=0.05$ and different $M$ with LISA, Taiji, and TianQin.
% The horizontal dashed lines represent the $3\sigma$ limit $33.3\%$.
}
\label{sigmaa}
\end{figure}

\begin{figure}
\centering
\includegraphics[width=0.9\columnwidth]{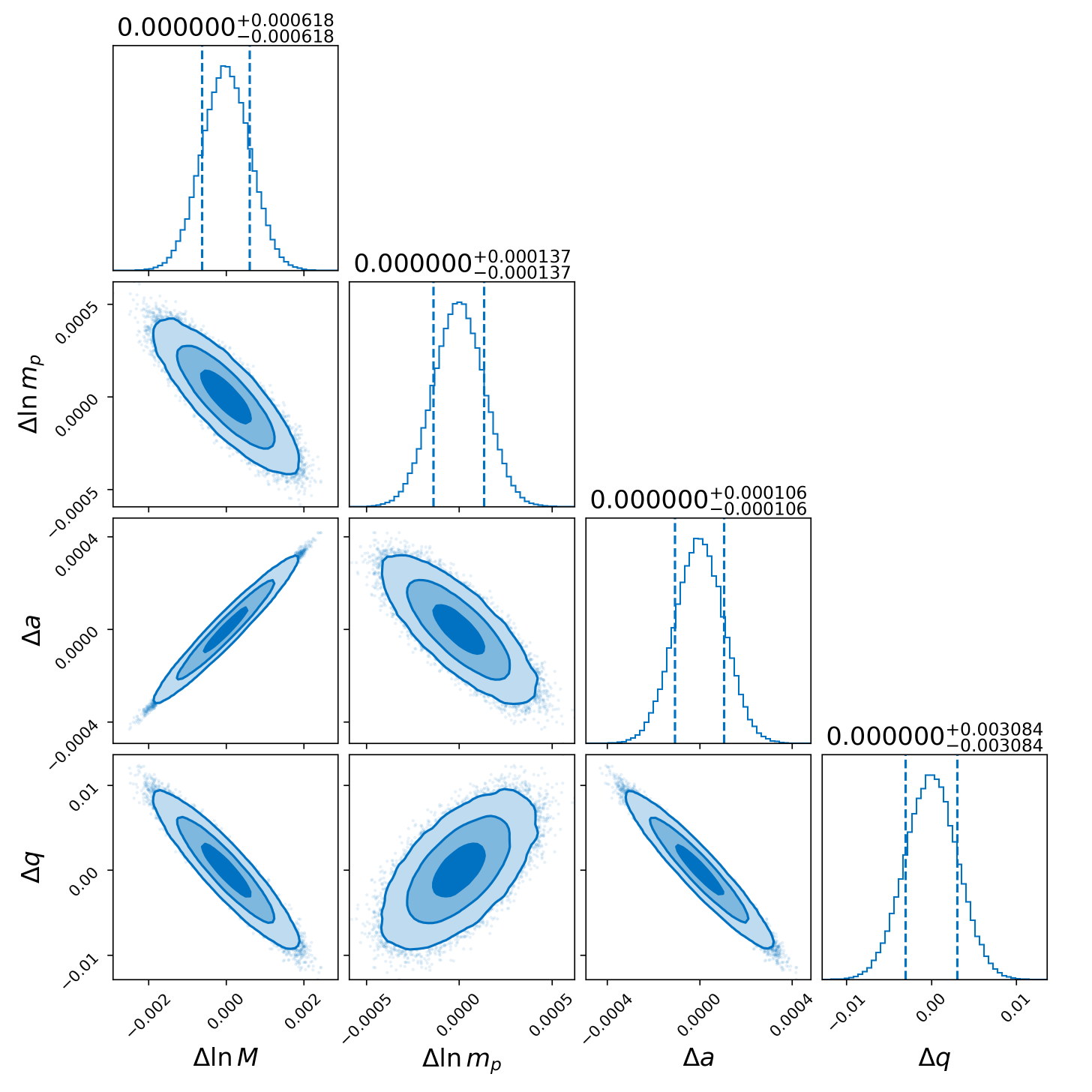}
\caption{Corner plot for the probability distribution of the source parameters $(\ln M,\ln m_p, a, q)$ with LISA, inferred after one-year observations of EMRIs with $q=0.05$ and $a=0.9$.
Vertical lines show the $1\sigma$ interval for the source parameter.
The contours correspond to the $68\%$, $95\%$, and $99\%$ probability confidence intervals.}
\label{corner09}
\end{figure}

\begin{figure}
\centering
\includegraphics[width=0.9\columnwidth]{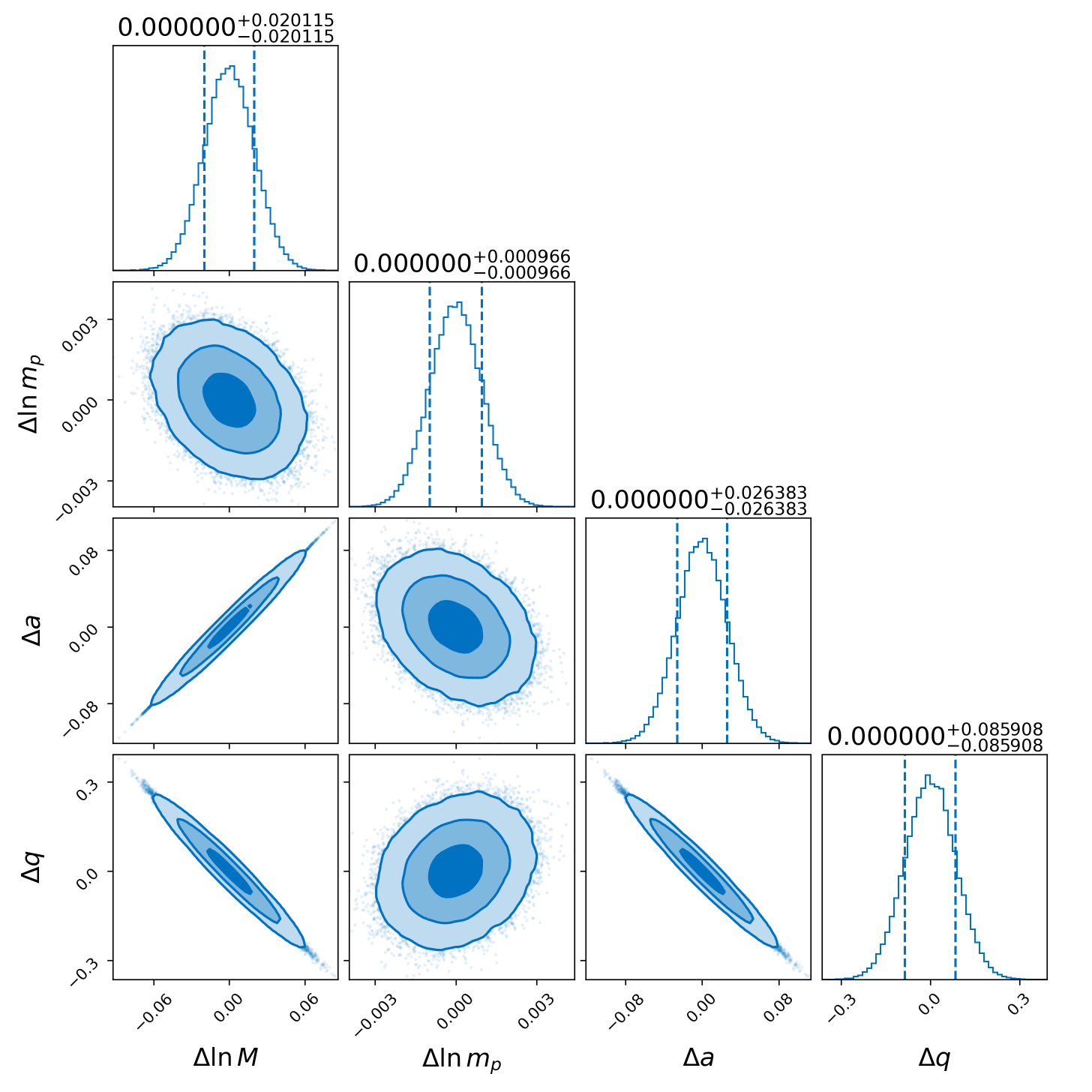}
\caption{Corner plot for the probability distribution of the source parameters $(\ln M,\ln m_p, a, q)$ with LISA, inferred after one-year observations of EMRIs with $q=0.05$ and $a=0$.
Vertical lines show the $1\sigma$ interval for the source parameter.
The contours correspond to the $68\%$, $95\%$, and $99\%$ probability confidence intervals.}
\label{corner0}
\end{figure}

\begin{figure}
\centering
\includegraphics[width=0.9\columnwidth]{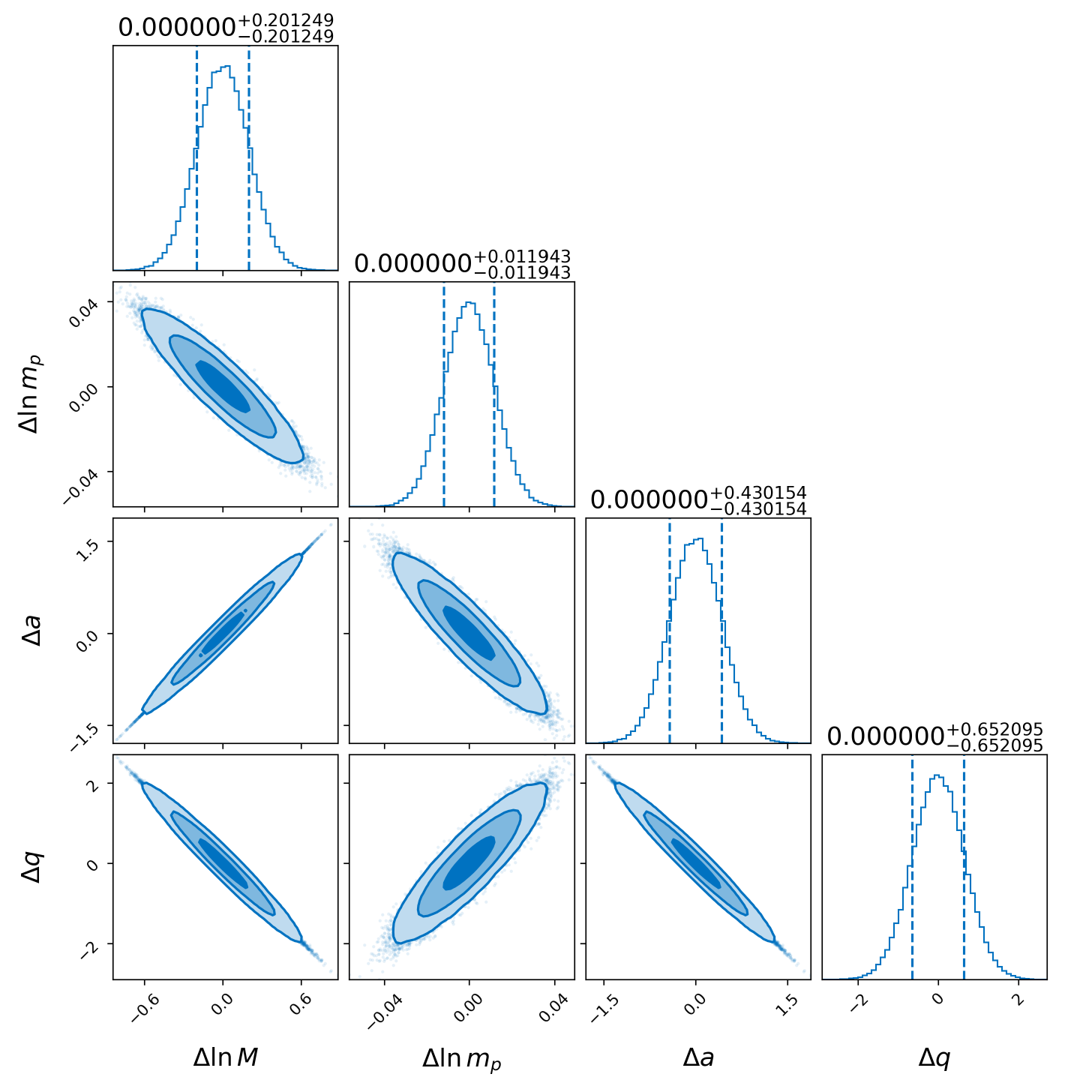}
\caption{Corner plot for the probability distribution of the source parameters $(\ln M,\ln m_p, a, q)$ with LISA, inferred after one-year observations of EMRIs with $q=0.05$ and $a=-0.9$.
Vertical lines show the $1\sigma$ interval for the source parameter.
The contours correspond to the $68\%$, $95\%$, and $99\%$ probability confidence intervals.}
\label{cornerm09}
\end{figure}

\begin{figure}
\centering
\includegraphics[width=0.9\columnwidth]{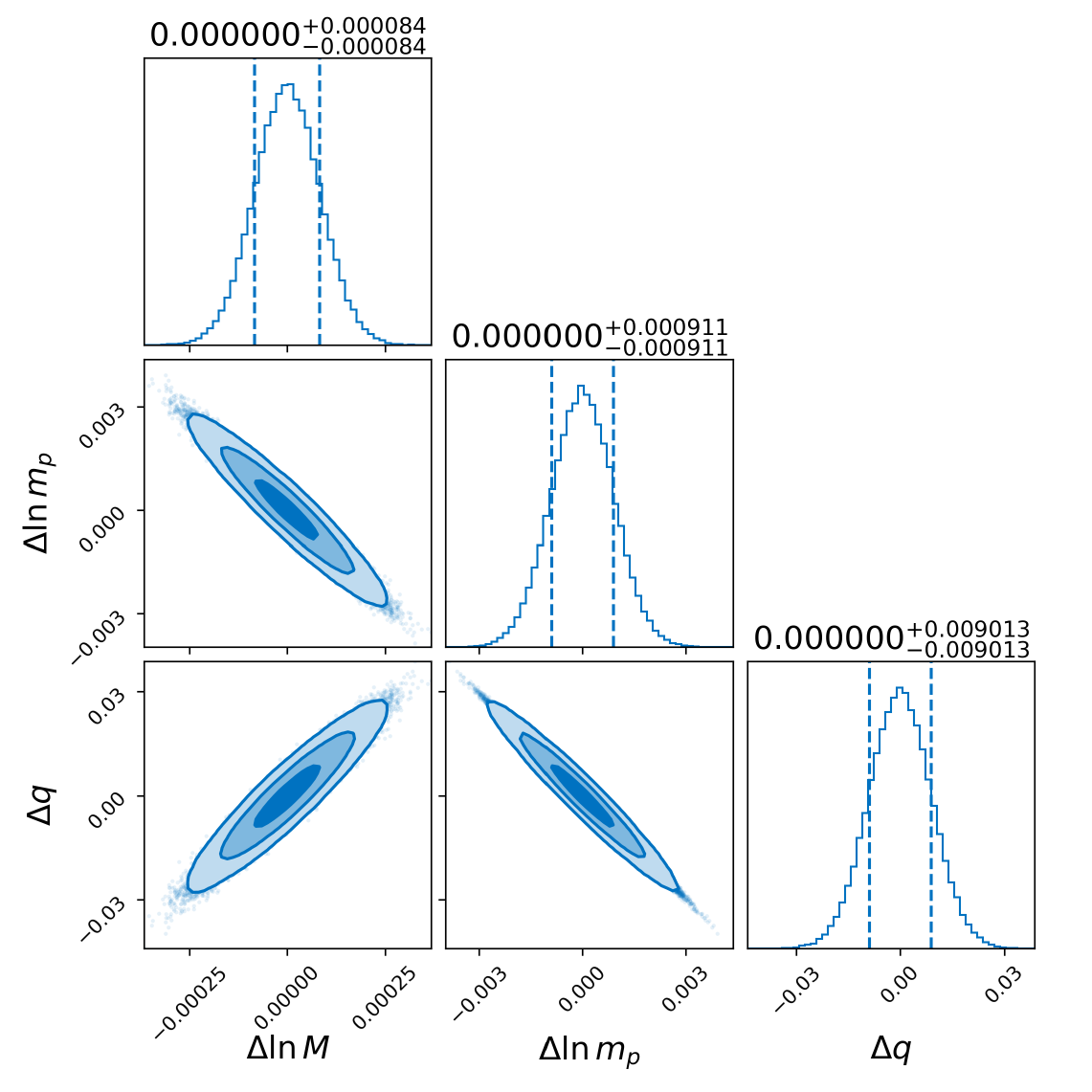}
\caption{Corner plot for the probability distribution of the source parameters $(\ln M,\ln m_p, q)$ with LISA, inferred after one-year observations of EMRIs with $q=0.05$ in the Schwarzschild BH background.
Vertical lines show the $1\sigma$ interval for the source parameter.
The contours correspond to the $68\%$, $95\%$, and $99\%$ probability confidence intervals.}
\label{corner0s}
\end{figure}

Figure \ref{sigmaa} shows the relative errors of the vector charge $q$ versus the spin $a$ with LISA, TianQin, and Taiji.
In general, the relative errors of the charge for the co-rotating orbit are better than those for the counter-rotating orbit.
We only consider the co-rotating orbit for simplicity.
The $1\sigma$ error for the charge decreases with the spin $a$.
Comparing the relative errors for Kerr BHs with spin $a=0.9$ and spin $a=0$,
we find that the spin of Kerr BHs can decrease the charge uncertainty by about one or two orders of magnitude, depending on the mass of the Kerr BH.
For EMRIs with $M=10^6~M_\odot$, $m_p=10~M_\odot$, $q=0.05$, and different $a$,
the corner plots for source parameters with LISA are shown in Figs. \ref{corner09}, \ref{corner0} and \ref{cornerm09}.
For comparison, we also show the corner plot for charged EMRIs in the Schwarzschild BH background in Fig. \ref{corner0s}.
For $a=(0.9,0,-0.9)$, the corresponding errors of charge $\sigma_q$ are $(0.0031,0.086,0.65)$, respectively.
As expected, $\sigma_q$ is smaller for co-rotating orbits and bigger for counter-rotating orbits.
For EMRIs in the Kerr background, the co-rotating orbit can better detect the vector charge.
It is interesting to note that the charge $q$ are anti-correlated with the mass $M$  and the spin $a$ of Kerr BHs,
and the correlations between $q$ and $M$, and $q$ and $m_p$ in the Kerr BH background
are opposite to those in the Schwarzschild BH background.

\section{Conclusions}
\label{sec6}
We study the energy emissions and GWs from EMRIs consisting of a small charged compact object with mass $m_p$ and the charge to mass ratio $q$ inspiraling into a Kerr BH with spin $a$.
We derive the formula for solving the inhomogeneous Teukolsky equation with a vector field and calculate the power emission due to the vector field in the Kerr background.
By using the difference between the number of GW cycles $\Delta\mathcal{N}$ accumulated by EMRIs with and without the vector charge in circular orbits over one year before the merger,
we may reveal the presence of a vector charge as small as $q\sim 0.007$ for Kerr BHs with $a=0.9$ and $M=10^6~M_\odot$, and $q\sim 0.01$ for Schwarzschild BHs with $M=10^6~M_\odot$.
The dephasing increases monotonically with the charge-to-mass ratio $q$, and it strongly depends on the mass of the Kerr BH such that lighter BHs have larger dephasing.

We also apply the faithfulness between GW signals with and without the vector charge
to discuss the detection of the vector charge $q$.
We find that positive larger spin of the Kerr BH can help us detect the vector charge easier.
We show that one-year observations of EMRIs with LISA may be able to reveal the presence of a vector charge as small as $q\sim 0.002$ for Kerr BHs with $a=0.9$ and $M=10^6~M_\odot$,
$q\sim 0.003$ for Schwarzschild BHs with $a=0$ and $M=10^6~M_\odot$, and $q\sim 0.004$ for Kerr BHs with $a=-0.9$ and $M=10^6~M_\odot$.

To determine vector charge more accurately and account for the degeneracy among parameters,
we calculate the FIM to estimate the errors of the vector charge $q$.
For EMRIs with $M=2\times10^5~M_{\odot}$ and $a=0.9$,
the vector charge $q$ can be constrained as small as $q\sim0.0049$ with TianQin, $q\sim0.0057$ with Taiji, and $q\sim 0.0082$ with LISA.
For EMRIs with $M=10^6~M_{\odot}$ and $a=0.9$,
the vector charge can be constrained as small as $q\sim 0.016$ with Taiji, $q\sim 0.021$ with LISA, and $q\sim 0.028$ with TianQin.
Since the sensitivity of TianQin is better than LISA and Taiji in the high-frequency bands,
the ability to detect the vector charge for TianQin is better than LISA and Taiji when the mass $M$ of the Kerr BH with $a=0.9$ is lighter than $\sim 2\times 10^5~M_{\odot}$.
For the mass $M$ of the Kerr BH with $a=0.9$ above $10^6~M_{\odot}$, LISA and Taiji are more likely to detect smaller vector charges.
The relative errors of the charge $q$ with Taiji are always smaller than LISA  because the sensitivity of Taiji is always better than LISA.
Due to the extra radiation of the vector field,
the charged particle starts further away from ISCO,
so the $1\sigma$ error for the charge decreases with the charge $q$.
As the spin $a$ of Kerr BHs increases, the ISCO becomes smaller,
the positive spin of Kerr BHs (co-rotating) can decrease the charge uncertainty by about one or two orders of magnitude, depending on the mass of the Kerr BH.
For EMRIs with $M=10^6~M_\odot$, $m_p=10~M_\odot$, $q=0.05$, and different $a=(0.9,0,-0.9)$, the corresponding errors of charge $\sigma_q$ with LISA are $(0.0031,0.086,0.65)$, respectively,
so co-rotating orbits can better detect the vector charge.
It is interesting to note that the charge $q$ are anti-correlated with the mass $M$  and the spin $a$ of Kerr BHs,
and the correlations between $q$ and $M$, and $q$ and $m_p$ in the Kerr BH background
are opposite to those in the Schwarzschild BH background.
In summary, the observations of EMRIs with a lighter and larger-spin Kerr BH can detect the vector charge easier.

\begin{acknowledgments}
This work makes use of the Black Hole Perturbation Toolkit package.
This research is supported in part by the National Key Research and Development Program of China under Grant No. 2020YFC2201504.
\end{acknowledgments}

\appendix
\section{Details for Calculating Energy Flux}
\subsection{Generalized Sasaki-Nakamura equation}\label{gsne}
In this appendix, we provide further technical details on the formalisms that we compute the vector flux.
The generalized Sasaki-Nakamura equation is given by \cite{Hughes:2000pf}
\begin{equation}
\frac{d^2X}{dr^{*2}} - {_{s}}F(r) \frac{dX}{dr^*} - {_{s}}U(r) X = 0,
\label{eq:SNeq}
\end{equation}
where $dr/dr^*=f(r)=\Delta/(r^2+a^2)$.
The coefficient ${_{s}}F(r)$ is
\begin{equation}
{_{s}}F(r)= \frac{\eta(r)_{,r}}{\eta(r)}\frac{\Delta}{r^2 + a^2},
\end{equation}
the function ${_{s}}U(r)$ is
\begin{eqnarray}
{_{s}}U(r) &=& {_{s}}U_1(r)\frac{\Delta}{(r^2 + a^2)^2} + {_{s}}G(r)^2 +
\frac{d{_{s}G(r)}}{dr}\frac{\Delta}{r^2 + a^2} -
\frac{\Delta{_{s}G(r)}{_{s}F_1(r)}}{r^2 + a^2},
\label{eq:gsn_potentials}
\end{eqnarray}
\begin{eqnarray}
{_{s}}F_1(r) &=& \frac{\eta(r)_{,r}}{\eta(r)},
\end{eqnarray}
\begin{eqnarray}
{_{s}}U_1(r) &=& V_T + \frac{1}{\beta\Delta^s}
\left[\left(2\alpha + \beta_{,r}\Delta^{s+1}\right)_{,r} - \frac{\eta(r)_{,r}}{\eta(r)}
\left(\alpha + \beta_{,r}\Delta^{s+1}\right)\right],
\label{eq:F1_and_U1}
\end{eqnarray}
\begin{equation}
{_{s}}G(r) = \frac{r\Delta}{(r^2 + a^2)^2} + \frac{s(r - 1)}{r^2 + a^2},
\label{eq:G_function}
\end{equation}
where ${}_{,r}$ denotes the derivative with respect to $r$ and
\begin{equation}
\eta(\hat{r}) = c_0 + \frac{c_1}{r} + \frac{c_2}{r^2} + \frac{c_3}{r^3} + \frac{c_4}{r^4}.
\end{equation}
The solutions of the Teukolsky and generalized Sasaki-Nakamura equations are related by
\begin{equation}
\label{eq:fromSNtoTeu}
\begin{split}
R_{lm\omega}(r)&=\frac{\left(
\alpha + \beta_{,r} \Delta^{s+1}\right) \chi -\beta\Delta^{s+1} \chi'}{\eta},   \\
\chi&=\frac{X}{{\sqrt{(r^2 + a^2)\Delta^s}}}.
\end{split}
\end{equation}
For $s=-1$,
\begin{equation}
\label{eq:em_eta_cofs}
\begin{split}
c_0 =& -\lambda_{lm\omega},\\
c_1 =& -2iam,\\
c_2 =& a^2(1 - 2\lambda_{lm\omega}),\\
c_3 =& -2a^2(1 + i a m),\\
c_4 =& a^4(1 - \lambda_{lm\omega}),
\end{split}
\end{equation}
and
\begin{equation}
\label{eq:em_alpha_beta}
\begin{split}
\alpha &= \frac{(r^2 + a^2)}{r^2}\sqrt{\Delta}\left[ - {\frac{r}{r^2 + a^2}} -
\frac{i K}{\Delta}\right],\\
\beta &= \frac{(r^2 + a^2)}{r^2}\sqrt{\Delta}.
\end{split}
\end{equation}

The generalized Sasaki-Nakamura equation admits two linearly independent solutions, $\Xin$ and $\Xup$, with the asymptotic behaviour
\begin{equation}
\Xin
\sim
\begin{cases}
e^{-i \kappa r^\ast} \quad  &r \to r_+  \\
A^{\textup{out}}_{l m \omega}  e^{i \omega r^\ast} + A^{\textup{in}}_{l m
\omega} e^{-i \omega r^\ast} \quad
&\hat{r} \to \infty  \label{eq:inBCSN}
\end{cases}\,,
\end{equation}
\begin{equation}
\Xup\sim
\begin{cases}
 C^{\textup{out}}_{l m \omega} e^{i \kappa r^{\ast}} + C^{\textup{in}}_{l m
\omega}e^{- i \kappa r^{\ast}} \quad \,  &r\to r_+   \label{eq:upBCSN} \\
e^{i \omega r^{\ast}}  \, \quad &r \to \infty
\end{cases}\,.
\end{equation}
With the above normalization of the solutions $\Xin$ $\Xup$, the arbitrary
constants $D^{\textup{tran}}$ and $B^{\textup{tran}}$ are determined as
\begin{equation}
D^{\textup{tran}}= \frac{2i\omega}{\lambda_{lm\omega}},
\end{equation}
\begin{equation}
B^{\textup{tran}}=\frac{i\sqrt{\sqrt{1-a^2}+1} }{4 \sqrt{2}  \left(2 \left(\sqrt{1-a^2}+1\right) \omega +i \sqrt{1-a^2}-a m\right)}.
\end{equation}
The numerical values of $\Xin$ ($\Xup$) are obtained by integrating
Eq.~\eqref{eq:SNeq} from $r_+$ (infinity) up to infinity ($r_+$)
using the boundary conditions~\ref{eq:inBCSN} (\ref{eq:upBCSN}).
We can derive the boundary conditions for the
homogeneous generalized Sasaki-Nakamura equation in terms of
explicit recursion relations which can be truncated at arbitrary order given by \cite{Piovano:2020zin}.
Then We transform  $\Xin, \Xup$ back to the Teukolsky solutions $\Rin, \Rup$ using Eq. \eqref{eq:fromSNtoTeu}. The amplitude
$B^{\textup{in}}_{\ell m \omega}$ can be obtained from the Wronskian $W$ at a given orbital separation.
\subsection{The source term}
In this subsection, we give the explicit expressions for the source terms to compute the amplitudes $Z_{\omega lm}^{\infty, H}$ for equatorial circular orbital configurations.
From Teukolsky's equation $(3.8)$ in \cite{Teukolsky:1973ha}, we get $J_2$
\begin{equation}
\begin{split}
J_2=&-\frac{3 \left(a^2+(r-2) r\right)}{2 (r-i a \cos (\theta ))^2 (r+i a \cos (\theta ))}J_{\bar{m}}\\
&-\frac{-\left(a^2+r^2\right)\partial_t J_{\bar{m}} +\left(a^2+(r-2) r\right)\partial_r J_{\bar{m}} -a \partial_\varphi J_{\bar{m}} }{2 (r-i a \cos (\theta )) (r+i a \cos (\theta ))}\\
&+\frac{i \sin (\theta ) \left(-2 a r-4 i a^2 \cos (\theta )\right) }{\sqrt{2} (r-i a \cos (\theta ))^2 (r+i a \cos (\theta ))}J_n\\
&+\frac{i \sin (\theta ) \left(a^2 \cos (2 \theta )+a^2+2 r^2\right) \left(a \partial_t J_n +\csc ^2(\theta ) \partial_\varphi J_n +i \csc (\theta ) \partial_\theta J_n\right)}{2 \sqrt{2} (r-i a \cos (\theta ))^2 (r+i a \cos (\theta ))},
\end{split}
\end{equation}
where $J_n=J^\mu n_{\mu}$ and $J_{\bar{m}}=J^\mu \bar{m}_\mu$.
The source term $T_{\omega lm}$ for $s=-1$ is
\begin{equation}
\begin{split}
T_{\omega lm}(r)&=\frac{1}{2\pi}\int dt d\Omega~ 4\pi \Sigma T {_{-1}}S^{m}_{\quad l}(\theta)e^{-im\varphi}e^{i\omega t}    \\
&=2\int dt d\Omega~e^{i\omega t-im\varphi}\frac{r^2+a^2\cos^2(\theta)}{(r-ia \cos(\theta))^{-2}} {_{-1}}S^{m}_{\quad l}(\theta) J_2.
\end{split}
\end{equation}
The amplitudes $Z_{\omega lm}^{\infty,H}$ in Eq.  \eqref{amplitudes} for $s=-1$ is
\begin{equation}
\int_{r_+}^{+\infty}\Delta^{-1}R^{\text{up},\text{in}}_{\omega lm}T_{\omega lm}dr=\delta(\omega-m \hat{\omega})\left[(I_{00}+I_{01})R^{\text{up},\text{in}}_{\omega lm}+I_1\frac{d R^{\text{up},\text{in}}_{\omega lm}}{dr}\right],
\end{equation}
where the coefficients are
\begin{equation}
I_{00}=\frac{\sqrt{2} \pi q \left(r~{_{-1}}S^{m}_{\quad l}~'\left(\frac{\pi }{2}\right)+i a ~{_{-1}}S^{m}_{\quad l}\left(\frac{\pi }{2}\right)\right)}{a \sqrt{r}+r^2},
\end{equation}
\begin{equation}
\begin{split}
I_{01}=&-\sqrt{2} \pi q~{_{-1}}S^{m}_{\quad l}\left(\frac{\pi}{2}\right)\left[    \frac{r \left(-r^2 \omega +(r-2) \left(m \sqrt{r}+i\right)\right)}{\left(a^2+(r-2) r\right) \left(a+r^{3/2}\right)}\right.   \\
&\left.+\frac{i a \left(a^2+a \sqrt{r} (1+i r \omega )+r \left(-i m \sqrt{r}-2 i r \omega +r-2\right)\right)}{\sqrt{r} \left(a^2+(r-2) r\right) \left(a+r^{3/2}\right)}\right],
\end{split}
\end{equation}
\begin{equation}
I_{1}=\frac{i \sqrt{2} \pi  q \sqrt{r}  \left(\sqrt{r}-a\right)}{a+r_0^{3/2}}~{_{-1}}S^{m}_{\quad l}\left(\frac{\pi}{2}\right).
\end{equation}

 % \bibliographystyle{jhep}
 % \bibliography{cEMRI}

\begin{thebibliography}{100}

\bibitem{Abbott:2016blz}
{\scshape LIGO Scientific and VIRGO Collaborations} collaboration,
  \emph{{Observation of Gravitational Waves from a Binary Black Hole Merger}},
  \href{https://doi.org/10.1103/PhysRevLett.116.061102}{\emph{Phys. Rev. Lett.}
  {\bfseries 116} (2016) 061102}
  [\href{https://arxiv.org/abs/1602.03837}{{\ttfamily 1602.03837}}].

\bibitem{TheLIGOScientific:2016agk}
{\scshape LIGO Scientific and VIRGO Collaborations} collaboration,
  \emph{{GW150914: The Advanced LIGO Detectors in the Era of First
  Discoveries}},
  \href{https://doi.org/10.1103/PhysRevLett.116.131103}{\emph{Phys. Rev. Lett.}
  {\bfseries 116} (2016) 131103}
  [\href{https://arxiv.org/abs/1602.03838}{{\ttfamily 1602.03838}}].

\bibitem{LIGOScientific:2018mvr}
{\scshape LIGO Scientific and VIRGO Collaborations} collaboration,
  \emph{{GWTC-1: A Gravitational-Wave Transient Catalog of Compact Binary
  Mergers Observed by LIGO and Virgo during the First and Second Observing
  Runs}}, \href{https://doi.org/10.1103/PhysRevX.9.031040}{\emph{Phys. Rev. X}
  {\bfseries 9} (2019) 031040}
  [\href{https://arxiv.org/abs/1811.12907}{{\ttfamily 1811.12907}}].

\bibitem{LIGOScientific:2020ibl}
{\scshape LIGO Scientific and VIRGO Collaborations} collaboration,
  \emph{{GWTC-2: Compact Binary Coalescences Observed by LIGO and Virgo During
  the First Half of the Third Observing Run}},
  \href{https://doi.org/10.1103/PhysRevX.11.021053}{\emph{Phys. Rev. X}
  {\bfseries 11} (2021) 021053}
  [\href{https://arxiv.org/abs/2010.14527}{{\ttfamily 2010.14527}}].

\bibitem{LIGOScientific:2021usb}
{\scshape LIGO Scientific and VIRGO Collaborations} collaboration,
  \emph{{GWTC-2.1: Deep Extended Catalog of Compact Binary Coalescences
  Observed by LIGO and Virgo During the First Half of the Third Observing
  Run}},  \href{https://arxiv.org/abs/2108.01045}{{\ttfamily 2108.01045}}.

\bibitem{LIGOScientific:2021djp}
{\scshape LIGO Scientific, VIRGO and KAGRA Collaborations} collaboration,
  \emph{{GWTC-3: Compact Binary Coalescences Observed by LIGO and Virgo During
  the Second Part of the Third Observing Run}},
  \href{https://arxiv.org/abs/2111.03606}{{\ttfamily 2111.03606}}.

\bibitem{LIGOScientific:2017vwq}
{\scshape LIGO Scientific and VIRGO Collaborations} collaboration,
  \emph{{GW170817: Observation of Gravitational Waves from a Binary Neutron
  Star Inspiral}},
  \href{https://doi.org/10.1103/PhysRevLett.119.161101}{\emph{Phys. Rev. Lett.}
  {\bfseries 119} (2017) 161101}
  [\href{https://arxiv.org/abs/1710.05832}{{\ttfamily 1710.05832}}].

\bibitem{LIGOScientific:2020aai}
{\scshape LIGO Scientific and VIRGO Collaborations} collaboration,
  \emph{{GW190425: Observation of a Compact Binary Coalescence with Total Mass
  $\sim 3.4 M_{\odot}$}},
  \href{https://doi.org/10.3847/2041-8213/ab75f5}{\emph{Astrophys. J. Lett.}
  {\bfseries 892} (2020) L3}
  [\href{https://arxiv.org/abs/2001.01761}{{\ttfamily 2001.01761}}].

\bibitem{LIGOScientific:2021qlt}
{\scshape LIGO Scientific, KAGRA and VIRGO Collaborations} collaboration,
  \emph{{Observation of Gravitational Waves from Two Neutron
  Star\textendash{}Black Hole Coalescences}},
  \href{https://doi.org/10.3847/2041-8213/ac082e}{\emph{Astrophys. J. Lett.}
  {\bfseries 915} (2021) L5}
  [\href{https://arxiv.org/abs/2106.15163}{{\ttfamily 2106.15163}}].

\bibitem{Danzmann:1997hm}
K.~Danzmann, \emph{{LISA: An ESA cornerstone mission for a gravitational wave
  observatory}}, \href{https://doi.org/10.1088/0264-9381/14/6/002}{\emph{Class.
  Quant. Grav.} {\bfseries 14} (1997) 1399}.

\bibitem{LISA:2017pwj}
{\scshape LISA} collaboration, \emph{{Laser Interferometer Space Antenna}},
  \href{https://arxiv.org/abs/1702.00786}{{\ttfamily 1702.00786}}.

\bibitem{Luo:2015ght}
{\scshape TianQin} collaboration, \emph{{TianQin: a space-borne gravitational
  wave detector}},
  \href{https://doi.org/10.1088/0264-9381/33/3/035010}{\emph{Class. Quant.
  Grav.} {\bfseries 33} (2016) 035010}
  [\href{https://arxiv.org/abs/1512.02076}{{\ttfamily 1512.02076}}].

\bibitem{Hu:2017mde}
W.-R.~Hu and Y.-L.~Wu, \emph{{The Taiji Program in Space for gravitational wave
  physics and the nature of gravity}},
  \href{https://doi.org/10.1093/nsr/nwx116}{\emph{Natl. Sci. Rev.} {\bfseries
  4} (2017) 685}.

\bibitem{Gong:2021gvw}
Y.~Gong, J.~Luo and B.~Wang, \emph{{Concepts and status of Chinese space
  gravitational wave detection projects}},
  \href{https://doi.org/10.1038/s41550-021-01480-3}{\emph{Nature Astron.}
  {\bfseries 5} (2021) 881} [\href{https://arxiv.org/abs/2109.07442}{{\ttfamily
  2109.07442}}].

\bibitem{TianQin:2020hid}
{\scshape TianQin} collaboration, \emph{{The TianQin project: current progress
  on science and technology}},
  \href{https://doi.org/10.1093/ptep/ptaa114}{\emph{PTEP} {\bfseries 2021}
  (2021) 05A107} [\href{https://arxiv.org/abs/2008.10332}{{\ttfamily
  2008.10332}}].

\bibitem{Ruan:2018tsw}
W.-H.~Ruan, Z.-K.~Guo, R.-G.~Cai and Y.-Z.~Zhang, \emph{{Taiji program:
  Gravitational-wave sources}},
  \href{https://doi.org/10.1142/S0217751X2050075X}{\emph{Int. J. Mod. Phys. A}
  {\bfseries 35} (2020) 2050075}
  [\href{https://arxiv.org/abs/1807.09495}{{\ttfamily 1807.09495}}].

\bibitem{LISA:2022yao}
{\scshape LISA} collaboration, \emph{{Astrophysics with the Laser
  Interferometer Space Antenna}},
  \href{https://doi.org/10.1007/s41114-022-00041-y}{\emph{Living Rev. Rel.}
  {\bfseries 26} (2023) 2} [\href{https://arxiv.org/abs/2203.06016}{{\ttfamily
  2203.06016}}].

\bibitem{LISA:2022kgy}
{\scshape LISA} collaboration, \emph{{New horizons for fundamental physics with
  LISA}}, \href{https://doi.org/10.1007/s41114-022-00036-9}{\emph{Living Rev.
  Rel.} {\bfseries 25} (2022) 4}
  [\href{https://arxiv.org/abs/2205.01597}{{\ttfamily 2205.01597}}].

\bibitem{Amaro-Seoane:2007osp}
P.~Amaro-Seoane, J.R.~Gair, M.~Freitag, M.~Coleman~Miller, I.~Mandel,
  C.J.~Cutler et~al., \emph{{Astrophysics, detection and science applications
  of intermediate- and extreme mass-ratio inspirals}},
  \href{https://doi.org/10.1088/0264-9381/24/17/R01}{\emph{Class. Quant. Grav.}
  {\bfseries 24} (2007) R113}
  [\href{https://arxiv.org/abs/astro-ph/0703495}{{\ttfamily
  astro-ph/0703495}}].

\bibitem{Babak:2017tow}
S.~Babak, J.~Gair, A.~Sesana, E.~Barausse, C.F.~Sopuerta, C.P.L.~Berry et~al.,
  \emph{{Science with the space-based interferometer LISA. V: Extreme
  mass-ratio inspirals}},
  \href{https://doi.org/10.1103/PhysRevD.95.103012}{\emph{Phys. Rev. D}
  {\bfseries 95} (2017) 103012}
  [\href{https://arxiv.org/abs/1703.09722}{{\ttfamily 1703.09722}}].

\bibitem{Barack:2003fp}
L.~Barack and C.~Cutler, \emph{{LISA capture sources: Approximate waveforms,
  signal-to-noise ratios, and parameter estimation accuracy}},
  \href{https://doi.org/10.1103/PhysRevD.69.082005}{\emph{Phys. Rev. D}
  {\bfseries 69} (2004) 082005}
  [\href{https://arxiv.org/abs/gr-qc/0310125}{{\ttfamily gr-qc/0310125}}].

\bibitem{Chua:2017ujo}
A.J.K.~Chua, C.J.~Moore and J.R.~Gair, \emph{{Augmented kludge waveforms for
  detecting extreme-mass-ratio inspirals}},
  \href{https://doi.org/10.1103/PhysRevD.96.044005}{\emph{Phys. Rev. D}
  {\bfseries 96} (2017) 044005}
  [\href{https://arxiv.org/abs/1705.04259}{{\ttfamily 1705.04259}}].

\bibitem{Gair:2010yu}
J.R.~Gair, C.~Tang and M.~Volonteri, \emph{{LISA extreme-mass-ratio inspiral
  events as probes of the black hole mass function}},
  \href{https://doi.org/10.1103/PhysRevD.81.104014}{\emph{Phys. Rev. D}
  {\bfseries 81} (2010) 104014}
  [\href{https://arxiv.org/abs/1004.1921}{{\ttfamily 1004.1921}}].

\bibitem{Holz:2005df}
D.E.~Holz and S.A.~Hughes, \emph{{Using gravitational-wave standard sirens}},
  \href{https://doi.org/10.1086/431341}{\emph{Astrophys. J.} {\bfseries 629}
  (2005) 15} [\href{https://arxiv.org/abs/astro-ph/0504616}{{\ttfamily
  astro-ph/0504616}}].

\bibitem{MacLeod:2007jd}
C.L.~MacLeod and C.J.~Hogan, \emph{{Precision of Hubble constant derived using
  black hole binary absolute distances and statistical redshift information}},
  \href{https://doi.org/10.1103/PhysRevD.77.043512}{\emph{Phys. Rev. D}
  {\bfseries 77} (2008) 043512}
  [\href{https://arxiv.org/abs/0712.0618}{{\ttfamily 0712.0618}}].

\bibitem{Laghi:2021pqk}
D.~Laghi, N.~Tamanini, W.~Del~Pozzo, A.~Sesana, J.~Gair, S.~Babak et~al.,
  \emph{{Gravitational-wave cosmology with extreme mass-ratio inspirals}},
  \href{https://doi.org/10.1093/mnras/stab2741}{\emph{Mon. Not. Roy. Astron.
  Soc.} {\bfseries 508} (2021) 4512}
  [\href{https://arxiv.org/abs/2102.01708}{{\ttfamily 2102.01708}}].

\bibitem{LISACosmologyWorkingGroup:2022jok}
{\scshape LISA Cosmology Working Group} collaboration, \emph{{Cosmology with
  the Laser Interferometer Space Antenna}},
  \href{https://arxiv.org/abs/2204.05434}{{\ttfamily 2204.05434}}.

\bibitem{eLISA:2013xep}
{\scshape eLISA} collaboration, \emph{{The Gravitational Universe}},
  \href{https://arxiv.org/abs/1305.5720}{{\ttfamily 1305.5720}}.

\bibitem{Eda:2013gg}
K.~Eda, Y.~Itoh, S.~Kuroyanagi and J.~Silk, \emph{{New Probe of Dark-Matter
  Properties: Gravitational Waves from an Intermediate-Mass Black Hole Embedded
  in a Dark-Matter Minispike}},
  \href{https://doi.org/10.1103/PhysRevLett.110.221101}{\emph{Phys. Rev. Lett.}
  {\bfseries 110} (2013) 221101}
  [\href{https://arxiv.org/abs/1301.5971}{{\ttfamily 1301.5971}}].

\bibitem{Eda:2014kra}
K.~Eda, Y.~Itoh, S.~Kuroyanagi and J.~Silk, \emph{{Gravitational waves as a
  probe of dark matter minispikes}},
  \href{https://doi.org/10.1103/PhysRevD.91.044045}{\emph{Phys. Rev. D}
  {\bfseries 91} (2015) 044045}
  [\href{https://arxiv.org/abs/1408.3534}{{\ttfamily 1408.3534}}].

\bibitem{Barausse:2014tra}
E.~Barausse, V.~Cardoso and P.~Pani, \emph{{Can environmental effects spoil
  precision gravitational-wave astrophysics?}},
  \href{https://doi.org/10.1103/PhysRevD.89.104059}{\emph{Phys. Rev. D}
  {\bfseries 89} (2014) 104059}
  [\href{https://arxiv.org/abs/1404.7149}{{\ttfamily 1404.7149}}].

\bibitem{Yue:2017iwc}
X.-J.~Yue and W.-B.~Han, \emph{{Gravitational waves with dark matter
  minispikes: the combined effect}},
  \href{https://doi.org/10.1103/PhysRevD.97.064003}{\emph{Phys. Rev. D}
  {\bfseries 97} (2018) 064003}
  [\href{https://arxiv.org/abs/1711.09706}{{\ttfamily 1711.09706}}].

\bibitem{Yue:2018vtk}
X.-J.~Yue, W.-B.~Han and X.~Chen, \emph{{Dark matter: an efficient catalyst for
  intermediate-mass-ratio-inspiral events}},
  \href{https://doi.org/10.3847/1538-4357/ab06f6}{\emph{Astrophys. J.}
  {\bfseries 874} (2019) 34}
  [\href{https://arxiv.org/abs/1802.03739}{{\ttfamily 1802.03739}}].

\bibitem{Berry:2019wgg}
C.P.L.~Berry, S.A.~Hughes, C.F.~Sopuerta, A.J.K.~Chua, A.~Heffernan,
  K.~Holley-Bockelmann et~al., \emph{{The unique potential of extreme
  mass-ratio inspirals for gravitational-wave astronomy}},
  \href{https://arxiv.org/abs/1903.03686}{{\ttfamily 1903.03686}}.

\bibitem{Hannuksela:2019vip}
O.A.~Hannuksela, K.C.Y.~Ng and T.G.F.~Li, \emph{{Extreme dark matter tests with
  extreme mass ratio inspirals}},
  \href{https://doi.org/10.1103/PhysRevD.102.103022}{\emph{Phys. Rev. D}
  {\bfseries 102} (2020) 103022}
  [\href{https://arxiv.org/abs/1906.11845}{{\ttfamily 1906.11845}}].

\bibitem{Destounis:2020kss}
K.~Destounis, A.G.~Suvorov and K.D.~Kokkotas, \emph{{Testing spacetime symmetry
  through gravitational waves from extreme-mass-ratio inspirals}},
  \href{https://doi.org/10.1103/PhysRevD.102.064041}{\emph{Phys. Rev. D}
  {\bfseries 102} (2020) 064041}
  [\href{https://arxiv.org/abs/2009.00028}{{\ttfamily 2009.00028}}].

\bibitem{Burton:2020wnj}
J.Y.J.~Burton and T.~Osburn, \emph{{Reissner-Nordstr\"om perturbation framework
  with gravitational wave applications}},
  \href{https://doi.org/10.1103/PhysRevD.102.104030}{\emph{Phys. Rev. D}
  {\bfseries 102} (2020) 104030}
  [\href{https://arxiv.org/abs/2010.12984}{{\ttfamily 2010.12984}}].

\bibitem{Torres:2020fye}
T.~Torres and S.R.~Dolan, \emph{{Electromagnetic self-force on a charged
  particle on Kerr spacetime: Equatorial circular orbits}},
  \href{https://doi.org/10.1103/PhysRevD.106.024024}{\emph{Phys. Rev. D}
  {\bfseries 106} (2022) 024024}
  [\href{https://arxiv.org/abs/2008.12703}{{\ttfamily 2008.12703}}].

\bibitem{Barausse:2020rsu}
E.~Barausse et~al., \emph{{Prospects for Fundamental Physics with LISA}},
  \href{https://doi.org/10.1007/s10714-020-02691-1}{\emph{Gen. Rel. Grav.}
  {\bfseries 52} (2020) 81} [\href{https://arxiv.org/abs/2001.09793}{{\ttfamily
  2001.09793}}].

\bibitem{Cardoso:2019rou}
V.~Cardoso and A.~Maselli, \emph{{Constraints on the astrophysical environment
  of binaries with gravitational-wave observations}},
  \href{https://doi.org/10.1051/0004-6361/202037654}{\emph{Astron. Astrophys.}
  {\bfseries 644} (2020) A147}
  [\href{https://arxiv.org/abs/1909.05870}{{\ttfamily 1909.05870}}].

\bibitem{Maselli:2020zgv}
A.~Maselli, N.~Franchini, L.~Gualtieri and T.P.~Sotiriou, \emph{{Detecting
  scalar fields with Extreme Mass Ratio Inspirals}},
  \href{https://doi.org/10.1103/PhysRevLett.125.141101}{\emph{Phys. Rev. Lett.}
  {\bfseries 125} (2020) 141101}
  [\href{https://arxiv.org/abs/2004.11895}{{\ttfamily 2004.11895}}].

\bibitem{Maselli:2021men}
A.~Maselli, N.~Franchini, L.~Gualtieri, T.P.~Sotiriou, S.~Barsanti and P.~Pani,
  \emph{{Detecting fundamental fields with LISA observations of gravitational
  waves from extreme mass-ratio inspirals}},
  \href{https://doi.org/10.1038/s41550-021-01589-5}{\emph{Nature Astron.}
  {\bfseries 6} (2022) 464} [\href{https://arxiv.org/abs/2106.11325}{{\ttfamily
  2106.11325}}].

\bibitem{Guo:2022euk}
H.~Guo, Y.~Liu, C.~Zhang, Y.~Gong, W.-L.~Qian and R.-H.~Yue, \emph{{Detection
  of scalar fields by extreme mass ratio inspirals with a Kerr black hole}},
  \href{https://doi.org/10.1103/PhysRevD.106.024047}{\emph{Phys. Rev. D}
  {\bfseries 106} (2022) 024047}
  [\href{https://arxiv.org/abs/2201.10748}{{\ttfamily 2201.10748}}].

\bibitem{Zhang:2022rfr}
C.~Zhang, Y.~Gong, D.~Liang and B.~Wang, \emph{{Gravitational waves from
  eccentric extreme mass-ratio inspirals as probes of scalar fields}},
  \href{https://arxiv.org/abs/2210.11121}{{\ttfamily 2210.11121}}.

\bibitem{Barsanti:2022ana}
S.~Barsanti, N.~Franchini, L.~Gualtieri, A.~Maselli and T.P.~Sotiriou,
  \emph{{Extreme mass-ratio inspirals as probes of scalar fields: Eccentric
  equatorial orbits around Kerr black holes}},
  \href{https://doi.org/10.1103/PhysRevD.106.044029}{\emph{Phys. Rev. D}
  {\bfseries 106} (2022) 044029}
  [\href{https://arxiv.org/abs/2203.05003}{{\ttfamily 2203.05003}}].

\bibitem{Barsanti:2022vvl}
S.~Barsanti, A.~Maselli, T.P.~Sotiriou and L.~Gualtieri, \emph{{Detecting
  massive scalar fields with Extreme Mass-Ratio Inspirals}},
  \href{https://arxiv.org/abs/2212.03888}{{\ttfamily 2212.03888}}.

\bibitem{Cardoso:2021wlq}
V.~Cardoso, K.~Destounis, F.~Duque, R.P.~Macedo and A.~Maselli, \emph{{Black
  holes in galaxies: Environmental impact on gravitational-wave generation and
  propagation}},
  \href{https://doi.org/10.1103/PhysRevD.105.L061501}{\emph{Phys. Rev. D}
  {\bfseries 105} (2022) L061501}
  [\href{https://arxiv.org/abs/2109.00005}{{\ttfamily 2109.00005}}].

\bibitem{Dai:2021olt}
N.~Dai, Y.~Gong, T.~Jiang and D.~Liang, \emph{{Intermediate mass-ratio
  inspirals with dark matter minispikes}},
  \href{https://doi.org/10.1103/PhysRevD.106.064003}{\emph{Phys. Rev. D}
  {\bfseries 106} (2022) 064003}
  [\href{https://arxiv.org/abs/2111.13514}{{\ttfamily 2111.13514}}].

\bibitem{Jiang:2021htl}
T.~Jiang, N.~Dai, Y.~Gong, D.~Liang and C.~Zhang, \emph{{Constraint on
  Brans-Dicke theory from intermediate/extreme mass ratio inspirals}},
  \href{https://doi.org/10.1088/1475-7516/2022/12/023}{\emph{JCAP} {\bfseries
  12} (2022) 023} [\href{https://arxiv.org/abs/2107.02700}{{\ttfamily
  2107.02700}}].

\bibitem{Zhang:2022hbt}
C.~Zhang and Y.~Gong, \emph{{Detecting electric charge with extreme mass ratio
  inspirals}}, \href{https://doi.org/10.1103/PhysRevD.105.124046}{\emph{Phys.
  Rev. D} {\bfseries 105} (2022) 124046}
  [\href{https://arxiv.org/abs/2204.08881}{{\ttfamily 2204.08881}}].

\bibitem{Gao:2022hho}
Q.~Gao, \emph{{Constraint on the mass of graviton with gravitational waves}},
  \href{https://doi.org/10.1007/s11433-022-1971-9}{\emph{Sci. China Phys. Mech.
  Astron.} {\bfseries 66} (2023) 220411}
  [\href{https://arxiv.org/abs/2206.02140}{{\ttfamily 2206.02140}}].

\bibitem{Gao:2022hsn}
Q.~Gao, Y.~You, Y.~Gong, C.~Zhang and C.~Zhang, \emph{{Testing alternative
  theories of gravity with space-based gravitational wave detectors}},
  \href{https://arxiv.org/abs/2212.03789}{{\ttfamily 2212.03789}}.

\bibitem{Destounis:2022obl}
K.~Destounis, A.~Kulathingal, K.D.~Kokkotas and G.O.~Papadopoulos,
  \emph{{Gravitational-wave imprints of compact and galactic-scale environments
  in extreme-mass-ratio binaries}},
  \href{https://arxiv.org/abs/2210.09357}{{\ttfamily 2210.09357}}.

\bibitem{Liang:2022gdk}
D.~Liang, R.~Xu, Z.-F.~Mai and L.~Shao, \emph{{Probing vector hair of black
  holes with extreme mass ratio inspirals}},
  \href{https://arxiv.org/abs/2212.09346}{{\ttfamily 2212.09346}}.

\bibitem{Cardoso:2020iji}
V.~Cardoso, C.F.B.~Macedo and R.~Vicente, \emph{{Eccentricity evolution of
  compact binaries and applications to gravitational-wave physics}},
  \href{https://doi.org/10.1103/PhysRevD.103.023015}{\emph{Phys. Rev. D}
  {\bfseries 103} (2021) 023015}
  [\href{https://arxiv.org/abs/2010.15151}{{\ttfamily 2010.15151}}].

\bibitem{Liu:2020vsy}
L.~Liu, O.~Christiansen, Z.-K.~Guo, R.-G.~Cai and S.P.~Kim,
  \emph{{Gravitational and electromagnetic radiation from binary black holes
  with electric and magnetic charges: Circular orbits on a cone}},
  \href{https://doi.org/10.1103/PhysRevD.102.103520}{\emph{Phys. Rev. D}
  {\bfseries 102} (2020) 103520}
  [\href{https://arxiv.org/abs/2008.02326}{{\ttfamily 2008.02326}}].

\bibitem{Liu:2020bag}
L.~Liu, O.~Christiansen, W.-H.~Ruan, Z.-K.~Guo, R.-G.~Cai and S.P.~Kim,
  \emph{{Gravitational and electromagnetic radiation from binary black holes
  with electric and magnetic charges: elliptical orbits on a cone}},
  \href{https://doi.org/10.1140/epjc/s10052-021-09849-4}{\emph{Eur. Phys. J. C}
  {\bfseries 81} (2021) 1048}
  [\href{https://arxiv.org/abs/2011.13586}{{\ttfamily 2011.13586}}].

\bibitem{Cardoso:2022whc}
V.~Cardoso, K.~Destounis, F.~Duque, R.~Panosso~Macedo and A.~Maselli,
  \emph{{Gravitational Waves from Extreme-Mass-Ratio Systems in Astrophysical
  Environments}},
  \href{https://doi.org/10.1103/PhysRevLett.129.241103}{\emph{Phys. Rev. Lett.}
  {\bfseries 129} (2022) 241103}
  [\href{https://arxiv.org/abs/2210.01133}{{\ttfamily 2210.01133}}].

\bibitem{Sotiriou:2013qea}
T.P.~Sotiriou and S.-Y.~Zhou, \emph{{Black hole hair in generalized
  scalar-tensor gravity}},
  \href{https://doi.org/10.1103/PhysRevLett.112.251102}{\emph{Phys. Rev. Lett.}
  {\bfseries 112} (2014) 251102}
  [\href{https://arxiv.org/abs/1312.3622}{{\ttfamily 1312.3622}}].

\bibitem{Silva:2017uqg}
H.O.~Silva, J.~Sakstein, L.~Gualtieri, T.P.~Sotiriou and E.~Berti,
  \emph{{Spontaneous scalarization of black holes and compact stars from a
  Gauss-Bonnet coupling}},
  \href{https://doi.org/10.1103/PhysRevLett.120.131104}{\emph{Phys. Rev. Lett.}
  {\bfseries 120} (2018) 131104}
  [\href{https://arxiv.org/abs/1711.02080}{{\ttfamily 1711.02080}}].

\bibitem{Doneva:2017bvd}
D.D.~Doneva and S.S.~Yazadjiev, \emph{{New Gauss-Bonnet Black Holes with
  Curvature-Induced Scalarization in Extended Scalar-Tensor Theories}},
  \href{https://doi.org/10.1103/PhysRevLett.120.131103}{\emph{Phys. Rev. Lett.}
  {\bfseries 120} (2018) 131103}
  [\href{https://arxiv.org/abs/1711.01187}{{\ttfamily 1711.01187}}].

\bibitem{Antoniou:2017acq}
G.~Antoniou, A.~Bakopoulos and P.~Kanti, \emph{{Evasion of No-Hair Theorems and
  Novel Black-Hole Solutions in Gauss-Bonnet Theories}},
  \href{https://doi.org/10.1103/PhysRevLett.120.131102}{\emph{Phys. Rev. Lett.}
  {\bfseries 120} (2018) 131102}
  [\href{https://arxiv.org/abs/1711.03390}{{\ttfamily 1711.03390}}].

\bibitem{Yunes:2011aa}
N.~Yunes, P.~Pani and V.~Cardoso, \emph{{Gravitational Waves from Quasicircular
  Extreme Mass-Ratio Inspirals as Probes of Scalar-Tensor Theories}},
  \href{https://doi.org/10.1103/PhysRevD.85.102003}{\emph{Phys. Rev. D}
  {\bfseries 85} (2012) 102003}
  [\href{https://arxiv.org/abs/1112.3351}{{\ttfamily 1112.3351}}].

\bibitem{Cardoso:2011xi}
V.~Cardoso, S.~Chakrabarti, P.~Pani, E.~Berti and L.~Gualtieri, \emph{{Floating
  and sinking: The Imprint of massive scalars around rotating black holes}},
  \href{https://doi.org/10.1103/PhysRevLett.107.241101}{\emph{Phys. Rev. Lett.}
  {\bfseries 107} (2011) 241101}
  [\href{https://arxiv.org/abs/1109.6021}{{\ttfamily 1109.6021}}].

\bibitem{Gibbons:1975kk}
G.W.~Gibbons, \emph{{Vacuum Polarization and the Spontaneous Loss of Charge by
  Black Holes}}, \href{https://doi.org/10.1007/BF01609829}{\emph{Commun. Math.
  Phys.} {\bfseries 44} (1975) 245}.

\bibitem{Eardley:1975kp}
D.M.~Eardley and W.H.~Press, \emph{{Astrophysical processes near black holes}},
  \href{https://doi.org/10.1146/annurev.aa.13.090175.002121}{\emph{Ann. Rev.
  Astron. Astrophys.} {\bfseries 13} (1975) 381}.

\bibitem{1982PhRvD..25.2509H}
R.S.~{Hanni}, \emph{{Limits on the charge of a collapsed object}},
  \href{https://doi.org/10.1103/PhysRevD.25.2509}{\emph{Phys. Rev. D}
  {\bfseries 25} (1982) 2509}.

\bibitem{1982Prama..18..385J}
P.S.~{Joshi} and J.V.~{Narlikar}, \emph{{Limits of the charge of a collapsed
  object.}}, \href{https://doi.org/10.1007/BF02848041}{\emph{Pramana}
  {\bfseries 18} (1982) 385}.

\bibitem{Gong:2019aqa}
Y.~Gong, Z.~Cao, H.~Gao and B.~Zhang, \emph{{On neutralization of charged black
  holes}}, \href{https://doi.org/10.1093/mnras/stz1904}{\emph{Mon. Not. Roy.
  Astron. Soc.} {\bfseries 488} (2019) 2722}
  [\href{https://arxiv.org/abs/1907.05239}{{\ttfamily 1907.05239}}].

\bibitem{Bekenstein:1971ej}
J.D.~Bekenstein, \emph{{Hydrostatic Equilibrium and Gravitational Collapse of
  Relativistic Charged Fluid Balls}},
  \href{https://doi.org/10.1103/PhysRevD.4.2185}{\emph{Phys. Rev. D} {\bfseries
  4} (1971) 2185}.

\bibitem{de1995relativistic}
F.~de~Felice, Y.~Yu and J.~Fang, \emph{Relativistic charged spheres},
  \href{https://doi.org/10.1093/mnras/277.1.L17}{\emph{Mon. Not. Roy. Astron.
  Soc.} {\bfseries 277} (1995) L17}.

\bibitem{deFelice:1999qp}
F.~de~Felice, S.-m.~Liu and Y.-q.~Yu, \emph{{Relativistic charged spheres. 2.
  Regularity and stability}},
  \href{https://doi.org/10.1088/0264-9381/16/8/307}{\emph{Class. Quant. Grav.}
  {\bfseries 16} (1999) 2669}
  [\href{https://arxiv.org/abs/gr-qc/9905099}{{\ttfamily gr-qc/9905099}}].

\bibitem{Ivanov:2002jy}
B.V.~Ivanov, \emph{{Static charged perfect fluid spheres in general
  relativity}}, \href{https://doi.org/10.1103/PhysRevD.65.104001}{\emph{Phys.
  Rev. D} {\bfseries 65} (2002) 104001}
  [\href{https://arxiv.org/abs/gr-qc/0203070}{{\ttfamily gr-qc/0203070}}].

\bibitem{Majumdar:1947eu}
S.D.~Majumdar, \emph{{A class of exact solutions of Einstein's field
  equations}}, \href{https://doi.org/10.1103/PhysRev.72.390}{\emph{Phys. Rev.}
  {\bfseries 72} (1947) 390}.

\bibitem{zhang_influence_1982}
J.L.~Zhang, W.Y.~Chau and T.Y.~Deng, \emph{The influence of a net charge on the
  critical mass of a neutron star},
  \href{https://doi.org/10.1007/BF00648990}{\emph{Astrophys. Space Sci.}
  {\bfseries 88} (1982) 81}.

\bibitem{Anninos:2001yb}
P.~Anninos and T.~Rothman, \emph{{Instability of extremal relativistic charged
  spheres}}, \href{https://doi.org/10.1103/PhysRevD.65.024003}{\emph{Phys. Rev.
  D} {\bfseries 65} (2002) 024003}
  [\href{https://arxiv.org/abs/gr-qc/0108082}{{\ttfamily gr-qc/0108082}}].

\bibitem{Bonnor:1975gba}
W.B.~Bonnor and S.B.P.~Wickramasuriya, \emph{{Are very large gravitational
  redshifts possible}}, {\emph{Mon. Not. Roy. Astron. Soc.} {\bfseries 170}
  (1975) 643}.

\bibitem{Ray:2003gt}
S.~Ray, A.L.~Espindola, M.~Malheiro, J.P.S.~Lemos and V.T.~Zanchin,
  \emph{{Electrically charged compact stars and formation of charged black
  holes}}, \href{https://doi.org/10.1103/PhysRevD.68.084004}{\emph{Phys. Rev.
  D} {\bfseries 68} (2003) 084004}
  [\href{https://arxiv.org/abs/astro-ph/0307262}{{\ttfamily
  astro-ph/0307262}}].

\bibitem{Wald:1974np}
R.M.~Wald, \emph{{Black hole in a uniform magnetic field}},
  \href{https://doi.org/10.1103/PhysRevD.10.1680}{\emph{Phys. Rev. D}
  {\bfseries 10} (1974) 1680}.

\bibitem{Cardoso:2016olt}
V.~Cardoso, C.F.B.~Macedo, P.~Pani and V.~Ferrari, \emph{{Black holes and
  gravitational waves in models of minicharged dark matter}},
  \href{https://doi.org/10.1088/1475-7516/2016/05/054}{\emph{JCAP} {\bfseries
  05} (2016) 054} [\href{https://arxiv.org/abs/1604.07845}{{\ttfamily
  1604.07845}}].

\bibitem{Bolton:2022hpt}
J.S.~Bolton, A.~Caputo, H.~Liu and M.~Viel, \emph{{Comparison of Low-Redshift
  Lyman-\ensuremath{\alpha} Forest Observations to Hydrodynamical Simulations
  with Dark Photon Dark Matter}},
  \href{https://doi.org/10.1103/PhysRevLett.129.211102}{\emph{Phys. Rev. Lett.}
  {\bfseries 129} (2022) 211102}
  [\href{https://arxiv.org/abs/2206.13520}{{\ttfamily 2206.13520}}].

\bibitem{Liu:2022wtq}
L.~Liu and S.P.~Kim, \emph{{Merger rate of charged black holes from the
  two-body dynamical capture}},
  \href{https://doi.org/10.1088/1475-7516/2022/03/059}{\emph{JCAP} {\bfseries
  03} (2022) 059} [\href{https://arxiv.org/abs/2201.02581}{{\ttfamily
  2201.02581}}].

\bibitem{Liu:2020cds}
L.~Liu, Z.-K.~Guo, R.-G.~Cai and S.P.~Kim, \emph{{Merger rate distribution of
  primordial black hole binaries with electric charges}},
  \href{https://doi.org/10.1103/PhysRevD.102.043508}{\emph{Phys. Rev. D}
  {\bfseries 102} (2020) 043508}
  [\href{https://arxiv.org/abs/2001.02984}{{\ttfamily 2001.02984}}].

\bibitem{Jai-akson:2017ldo}
P.~Jai-akson, A.~Chatrabhuti, O.~Evnin and L.~Lehner, \emph{{Black hole merger
  estimates in Einstein-Maxwell and Einstein-Maxwell-dilaton gravity}},
  \href{https://doi.org/10.1103/PhysRevD.96.044031}{\emph{Phys. Rev. D}
  {\bfseries 96} (2017) 044031}
  [\href{https://arxiv.org/abs/1706.06519}{{\ttfamily 1706.06519}}].

\bibitem{Christiansen:2020pnv}
O.~Christiansen, J.~Beltr\'an~Jim\'enez and D.F.~Mota, \emph{{Charged Black
  Hole Mergers: Orbit Circularisation and Chirp Mass Bias}},
  \href{https://doi.org/10.1088/1361-6382/abdaf5}{\emph{Class. Quant. Grav.}
  {\bfseries 38} (2021) 075017}
  [\href{https://arxiv.org/abs/2003.11452}{{\ttfamily 2003.11452}}].

\bibitem{Bozzola:2020mjx}
G.~Bozzola and V.~Paschalidis, \emph{{General Relativistic Simulations of the
  Quasicircular Inspiral and Merger of Charged Black Holes: GW150914 and
  Fundamental Physics Implications}},
  \href{https://doi.org/10.1103/PhysRevLett.126.041103}{\emph{Phys. Rev. Lett.}
  {\bfseries 126} (2021) 041103}
  [\href{https://arxiv.org/abs/2006.15764}{{\ttfamily 2006.15764}}].

\bibitem{Zi:2022hcc}
T.~Zi, Z.~Zhou, H.-T.~Wang, P.-C.~Li, J.-d.~Zhang and B.~Chen, \emph{{Analytic
  kludge waveforms for extreme-mass-ratio inspirals of a charged object around
  a Kerr-Newman black hole}},
  \href{https://doi.org/10.1103/PhysRevD.107.023005}{\emph{Phys. Rev. D}
  {\bfseries 107} (2023) 023005}
  [\href{https://arxiv.org/abs/2205.00425}{{\ttfamily 2205.00425}}].

\bibitem{Teukolsky:1973ha}
S.A.~Teukolsky, \emph{{Perturbations of a rotating black hole. 1. Fundamental
  equations for gravitational electromagnetic and neutrino field
  perturbations}}, \href{https://doi.org/10.1086/152444}{\emph{Astrophys. J.}
  {\bfseries 185} (1973) 635}.

\bibitem{Press:1973zz}
W.H.~Press and S.A.~Teukolsky, \emph{{Perturbations of a Rotating Black Hole.
  II. Dynamical Stability of the Kerr Metric}},
  \href{https://doi.org/10.1086/152445}{\emph{Astrophys. J.} {\bfseries 185}
  (1973) 649}.

\bibitem{Teukolsky:1974yv}
S.A.~Teukolsky and W.H.~Press, \emph{{Perturbations of a rotating black hole.
  III - Interaction of the hole with gravitational and electromagnet ic
  radiation}}, \href{https://doi.org/10.1086/153180}{\emph{Astrophys. J.}
  {\bfseries 193} (1974) 443}.

\bibitem{Newman:1966ub}
E.T.~Newman and R.~Penrose, \emph{{Note on the Bondi-Metzner-Sachs group}},
  \href{https://doi.org/10.1063/1.1931221}{\emph{J. Math. Phys.} {\bfseries 7}
  (1966) 863}.

\bibitem{Goldberg:1966uu}
J.N.~Goldberg, A.J.~MacFarlane, E.T.~Newman, F.~Rohrlich and E.C.G.~Sudarshan,
  \emph{{Spin-s spherical harmonics and $\eth$}},
  \href{https://doi.org/10.1063/1.1705135}{\emph{J. Math. Phys.} {\bfseries 8}
  (1967) 2155}.

\bibitem{BHPToolkit}
``{Black Hole Perturbation Toolkit}.''
  (\href{http://bhptoolkit.org/}{bhptoolkit.org}).

\bibitem{Detweiler:1978ge}
S.L.~Detweiler, \emph{{Black Holes and Gravitational Waves. I. Circular Orbits
  About a Rotating Hole}},
  \href{https://doi.org/10.1086/156529}{\emph{Astrophys. J.} {\bfseries 225}
  (1978) 687}.

\bibitem{Hughes:1999bq}
S.A.~Hughes, \emph{{The Evolution of circular, nonequatorial orbits of Kerr
  black holes due to gravitational wave emission}},
  \href{https://doi.org/10.1103/PhysRevD.65.069902}{\emph{Phys. Rev. D}
  {\bfseries 61} (2000) 084004}
  [\href{https://arxiv.org/abs/gr-qc/9910091}{{\ttfamily gr-qc/9910091}}].

\bibitem{Jefremov:2015gza}
P.I.~Jefremov, O.Y.~Tsupko and G.S.~Bisnovatyi-Kogan, \emph{{Innermost stable
  circular orbits of spinning test particles in Schwarzschild and Kerr
  space-times}}, \href{https://doi.org/10.1103/PhysRevD.91.124030}{\emph{Phys.
  Rev. D} {\bfseries 91} (2015) 124030}
  [\href{https://arxiv.org/abs/1503.07060}{{\ttfamily 1503.07060}}].

\bibitem{Berti:2004bd}
E.~Berti, A.~Buonanno and C.M.~Will, \emph{{Estimating spinning binary
  parameters and testing alternative theories of gravity with LISA}},
  \href{https://doi.org/10.1103/PhysRevD.71.084025}{\emph{Phys. Rev. D}
  {\bfseries 71} (2005) 084025}
  [\href{https://arxiv.org/abs/gr-qc/0411129}{{\ttfamily gr-qc/0411129}}].

\bibitem{Lindblom:2008cm}
L.~Lindblom, B.J.~Owen and D.A.~Brown, \emph{{Model Waveform Accuracy Standards
  for Gravitational Wave Data Analysis}},
  \href{https://doi.org/10.1103/PhysRevD.78.124020}{\emph{Phys. Rev. D}
  {\bfseries 78} (2008) 124020}
  [\href{https://arxiv.org/abs/0809.3844}{{\ttfamily 0809.3844}}].

\bibitem{Chatziioannou:2017tdw}
K.~Chatziioannou, A.~Klein, N.~Yunes and N.~Cornish, \emph{{Constructing
  Gravitational Waves from Generic Spin-Precessing Compact Binary Inspirals}},
  \href{https://doi.org/10.1103/PhysRevD.95.104004}{\emph{Phys. Rev. D}
  {\bfseries 95} (2017) 104004}
  [\href{https://arxiv.org/abs/1703.03967}{{\ttfamily 1703.03967}}].

\bibitem{Hughes:2000pf}
S.A.~Hughes, \emph{{Computing radiation from Kerr black holes: Generalization
  of the Sasaki-Nakamura equation}},
  \href{https://doi.org/10.1103/PhysRevD.62.044029}{\emph{Phys. Rev. D}
  {\bfseries 62} (2000) 044029}
  [\href{https://arxiv.org/abs/gr-qc/0002043}{{\ttfamily gr-qc/0002043}}].

\bibitem{Piovano:2020zin}
G.A.~Piovano, A.~Maselli and P.~Pani, \emph{{Extreme mass ratio inspirals with
  spinning secondary: a detailed study of equatorial circular motion}},
  \href{https://doi.org/10.1103/PhysRevD.102.024041}{\emph{Phys. Rev. D}
  {\bfseries 102} (2020) 024041}
  [\href{https://arxiv.org/abs/2004.02654}{{\ttfamily 2004.02654}}].

\end{thebibliography}

\providecommand{\href}[2]{#2}\begingroup\raggedright\endgroup

\end{document}